\documentclass[a4paper,11pt]{article}
\usepackage{jcappub}
\usepackage{bm}
\usepackage{amsmath}
\usepackage{epsfig}

\begin{document}
\hyphenation{GW}
\def\prd{Phys. Rev. D}
\def\prl{Phys. Rev. Lett.}
\def\apj{Astrophys. J.}
\def\apjl{Astrophys. J. Lett.}
\def\apjs{Astrophys. J.Suppl.}
\def\mnras{Mon. Not. R. Astr. Soc.}
\def\aap{Astr. Astrophys.}
\def\aaps{Astr. Astrophys. Suppl}
\def\physrep{Phys. Rep.}
\def\aj{Astr. J.}
\def\nat{Nature}
\def\jcap{JCAP}
\def\ijmpd{Int. J. Mod. Phys. D}
\newcommand{\bi}[1]{\mbox{\boldmath $#1$}}
\newcommand{\nc}{\newcommand}
\newcommand{\beq}{\begin{equation}}
\newcommand{\eeq}{\end{equation}}
\newcommand{\be}{\begin{eqnarray}}
\newcommand{\ee}{\end{eqnarray}}
\newcommand{\num}{\nu_\mu}
\newcommand{\nue}{\nu_e}
\newcommand{\nut}{\nu_\tau}
\newcommand{\nus}{\nu_s}
\newcommand{\mnus}{M_s}
\newcommand{\taus}{\tau_{\nu_s}}
\newcommand{\nnt}{n_{\nu_\tau}}
\newcommand{\rnt}{\rho_{\nu_\tau}}
\newcommand{\mnt}{m_{\nu_\tau}}
\newcommand{\tnt}{\tau_{\nu_\tau}}
\newcommand{\rar}{\rightarrow}
\newcommand{\lar}{\leftarrow}
\newcommand{\lrar}{\leftrightarrow}
\newcommand{\dm}{\delta m^2}
\newcommand{\mpl}{m_{Pl}}
\newcommand{\mbh}{M_{BH}}
\newcommand{\nbh}{n_{BH}}
\newcommand{\crit}{{\rm crit}}
\newcommand{\ini}{{\rm in}}
\newcommand{\cmb}{{\rm cmb}}
\newcommand{\rec}{{\rm rec}}

\newcommand{\Odm}{\Omega_{\rm dm}}
\newcommand{\Ob}{\Omega_{\rm b}}
\newcommand{\Om}{\Omega_{\rm m}}
\newcommand{\nb}{n_{\rm b}}
\def\simlt{\lesssim}
\def\simgt{\gtrsim}
\def\Cl{C_{\ell}}
\def\out{{\rm out}}
\def\in{{\rm in}}
\def\mean{{\rm mean}}
\def\zrec{z_{\rm rec}}
\def\zreio{z_{\rm reion}}
\def\wmap{{\it WMAP }}
\def\planck{{\it Planck }}

\title{Understanding the LIGO GW150914 event}
%\author{ Pavel Naselsky, Andrew D. Jackson and Hao Liu}

\author[a]{Pavel Naselsky}
\author[a]{Andrew D. Jackson}
\author[a,b]{ Hao Liu} %\email[mail to: ]{liuhao@nbi.dk}
\affiliation[a]{Niels Bohr Institute \& Discovery Center, Blegdamsvej 17, DK-2100 Copenhagen, Denmark}
\affiliation[b]{Key laboratory of Particle and Astrophysics, Institute of High Energy Physics, CAS, China}

\emailAdd{liuhao@nbi.dk}

%\begin{abstract}
\abstract{
We present a simplified method for the extraction of meaningful signals from Hanford and Livingston 32 second data for the GW150914 event made publicly available by the LIGO collaboration and demonstrate its ability to reproduce the LIGO collaboration's own results quantitatively given the assumption that all narrow peaks in the power spectrum are a consequence of physically uninteresting signals and can be removed. After the clipping of these peaks and return to the time domain, the GW150914 event is readily distinguished from broadband background noise. This simple technique allows us to identify the GW150914 event without any assumption regarding its physical origin and with minimal assumptions regarding its shape. We also confirm that the LIGO GW150914 event is uniquely correlated in the Hanford and Livingston detectors for the full 4096 second
data at the level of $6-7\,\sigma$ with a temporal displacement of $\tau=6.9 \pm 0.4\,$ms. We have also identified a few events
that are morphologically close to GW150914 but less strongly cross correlated with it.
}
%\end{abstract}
\maketitle
\section{Introduction}
 The recent announcement by the LIGO Collaboration \citep{LIGO PRL} of the first observation of a gravitational wave (GW150914) and its theoretical interpretation as the merger of two massive black holes has generated interest and excitement well beyond the borders of the scientific community. Quite aside from the evident scientific importance of these results, the ability to detect strains as small $10^{-21}$ is a truly extraordinary scientific and engineering triumph. While such sensitivity was essential in detecting gravitational waves, it also results in the observability of noise from a wide range of sources. The LIGO power spectrum includes narrowband effects, e.g. from mechanical resonances and power sources (60 Hz and harmonics), and more troublesome broadband noise, e.g. from seismic, thermal and quantum effects. The net effect of all noises sources (including the fact that they are time dependent) is roughly three orders of magnitude larger than the amplitude of the GW150914 event and thus must be understood in considerable detail. This matter has been treated with care in \citep{LIGO Tech}.

However, given the significance LIGO GW150914 event, it is important to demonstrate the robustness of its detection by using
methods that are based on minimal assumptions regarding the properties of the signal and are without special assumptions
regarding the physical origins of both sources and noise. The aim of the present paper is to suggest such a method and to apply it to the analysis of GW150914.

The task is to discover temporally localized identical signals that appear in the LIGO Hanford Observatory detector (H) and the LIGO Livingston Observatory detector (L) with a time delay of $\pm\tau=0-10$\,ms with a possible sign inversion
of the records depending on the direction of the signal. We first construct the power spectra of H32 and L32. Unwanted
low ($\nu \le 30$\,Hz) and high ($\nu \ge 300$\,Hz) frequency information is filtered out, and narrow peaks are clipped.
The filtered time records are then constructed, and the Hanford record is shifted by time $\pm \tau = 0-10$\,ms. We then
consider the 160 patches of length 200\,ms contained in the 32 second H32 (shifted) and L32 records and calculate the H-L
cross-correlation coefficients for all patches. The H-L patch with the largest cross-correlation is precisely the GW150914 event with $\tau=6.9$\,ms and with inversion of the H32 data. The cross-correlation for all 200 ms records (inverted and shifted
by $\tau=6.9$\,ms) is then considered and confirms the uniqueness of GW150914 in both the 32\,s and 4096\,s records. Ordering
the cross-correlation coefficients, we can determine the distribution function and classify the morphological similarity of all
patches to the GW150914 patch. We stress that our method is blind with respect to any assumption about the astrophysical origin of the GW150914 signal and that it is sensitive to the morphology (phases) of the signal at all 200\,ms records.

\section{Description of the time ordered data analysis}
The LIGO Collaboration has made available two records of the GW event of September 14, 2015 ~\citep{LigoData2016,LigoData}.
These records, of length 32\,s and 4096\,s, describe the strains measured at the LIGO Hanford ($S_H$) and Livingston ($S_L$) observatories as a function of time~\citep{LigoData}. We consider first the H32 and L32\,s records. These records can be written as
\begin{eqnarray}
&&S_L(t)=G(t)+n_L(t)+F_L(t), \nonumber\\
 &&S_H(t)=(\pm)G(t\pm\tau)+n_H(t)+F_H(t),
\label{eq2}
\end{eqnarray}
where $G(t)$ is an event observed at detector L and observed (and possibly inverted) at detector H at a time shifted by
$\pm\tau$. Here, $n_{H,L}(t)$ are the H and L components of broadband noise, and the $F_{H,L}$ accounts for narrowband
systematic effects that can be time-dependent. In the following, we will assume that the amplitude of the $G$-signal is
greater than the noise components, $|G| > |n_{H,L}|$. The primary task is to reduce ``foreground'' components
$F_{H,L}$ to a level comparable to that of $n_{H,L}$. The full description of how this reduction is realized by LIGO is given in \citep{LIGO Tech}. Independent confirmation assumes that at least the major properties of the GW150914 event can be recovered by a blind technique based on very general assumptions regarding foreground effects and the properties of the signal under investigation (i.e., without the use of templates for the sources). Our processing of the Time Ordered Data (TOD) consists of the
following steps:\\
\textbullet ~We start our analysis using the 32 second H32 and L32 records.\\
\textbullet ~Treating data from the two sites separately, we construct the fast Fourier transform (FFT) of these records to obtain a power spectrum which contains narrow peaks and a smooth background of broadband noise. The narrow peaks are then ``clipped'' down to zero power. Low ($\nu \le 30$\,Hz ) and high ($\nu\ge 300$\,Hz) frequency filters were implemented following the recommendations of LIGO. \\
\textbullet ~ The clipped Livingston power spectrum is inverse Fourier transformed to obtain the modified record, $\overline{S}_L (t)$. The clipped Hanford power spectrum is also inverse Fourier transformed to $\overline{S}_H(t)$. We then follow~\citep{LIGO PRL} by inverting this function and displacing it by a time $\tau \approx 6.9$\,ms to obtained the modified
record for the Hanford observatory, $\overline{S}_H(t,\tau)$. \\
 \textbullet ~ For a given choice of $\tau$, we divide the records into shorter patches of length 200\,ms. For each pair of H-L patches we calculate the Pearson cross-correlation coefficient, $ C(\tau) $, defined as
\begin{eqnarray}
&&C({\tau}) = \frac{1}{N}\sum_j \, \overline{s}_H (t_j) \overline{s}_L (t_j, \tau) ,\nonumber\\
&&N= \left( \sum_i \, \overline{s}^2_H (t_i) \
\sum_k \, \overline{s}^2_L (t_k, \tau) \right)^{1/2} \, ,\nonumber\\
&&\overline{s}_{H,L}=\overline{S}_{H,L}-\langle \overline{S}_{H,L}\rangle \,.
\label{eq1}
\end{eqnarray}
Here, the sums extend over all strain measurements in the given time interval. The desired signal is identified
as being contained in the patch with the maximum value of $C(\tau)$. \footnote{The displacement time, $\tau$, in the interval $[0,10]$\,ms was varied and the analysis repeated with the aim of maximizing $C(\tau)$. We have confirmed that the peak of $C(\tau)$ is maximized when $\tau=6.9 \pm 0.4$\,ms.}

\section{ Cross-correlation analysis for H32 and L32 records.}
\begin{figure}[!htb]
 \begin{center}
\hbox{
 \centerline{\includegraphics[scale=0.13]{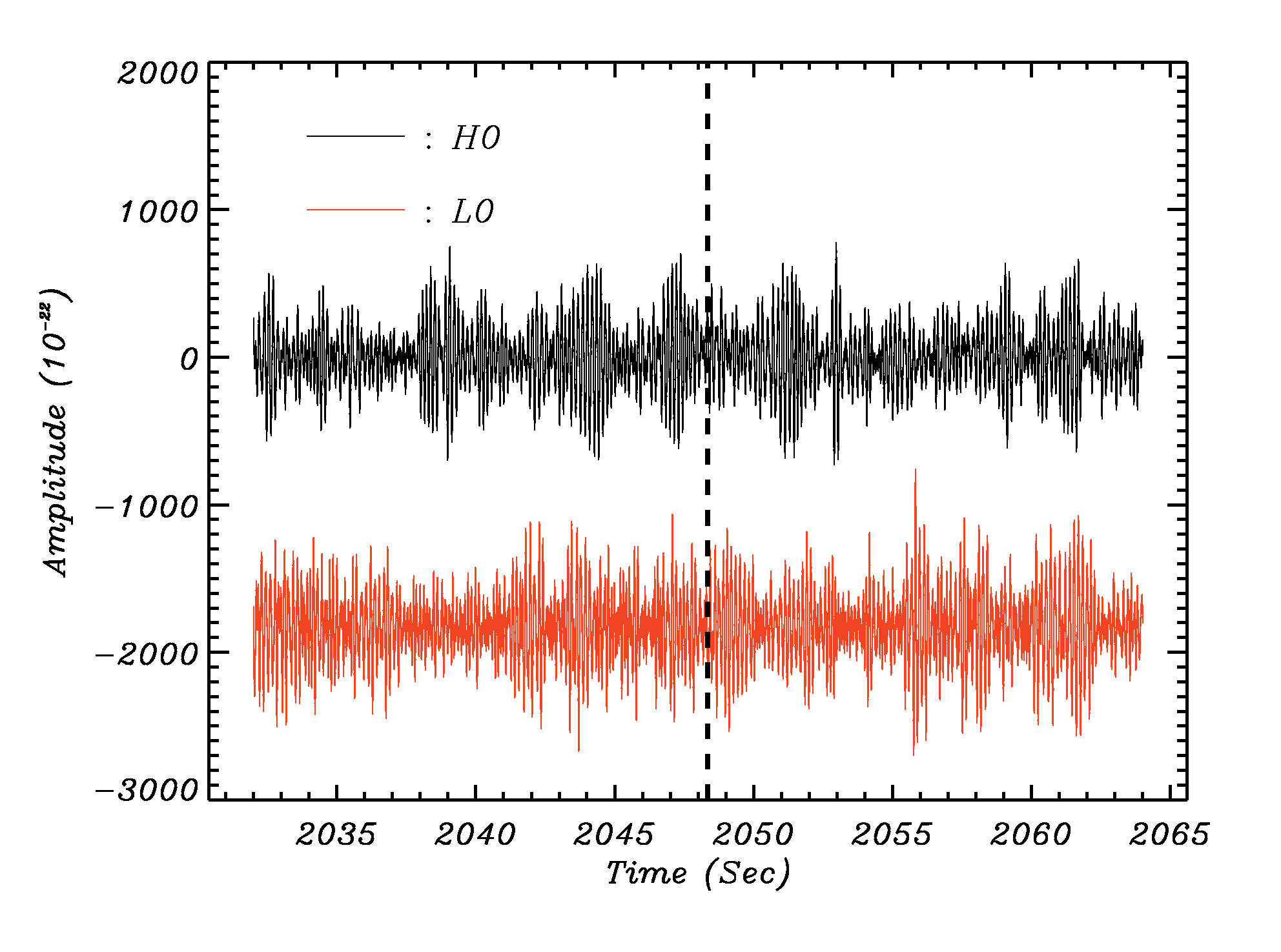}}}
 \caption{The H32 and L32 records before cleaning . The black line corresponds to the Hanford
 32\,s record, and the red line is for the Livingston 32\,s record before clipping. The vertical dashed
 line indicates the position of GW150914 event. }
 \label{fig1}
 \end{center}
\end{figure}

Here, we will illustrate the application of the method described above to the H32 and L32 records. Fig.~\ref{fig1} shows the original data from LIGO before any clipping, displacements and inversions.
Note the strong modulations of the data by sinusoidal modes that directly indicate the existence of large amplitude peaks in the power spectrum as discussed in \citep{LIGO PRL}.
\begin{figure*}[!htb]
 \begin{center}
\hbox{
 \centerline{\includegraphics[scale=0.14]{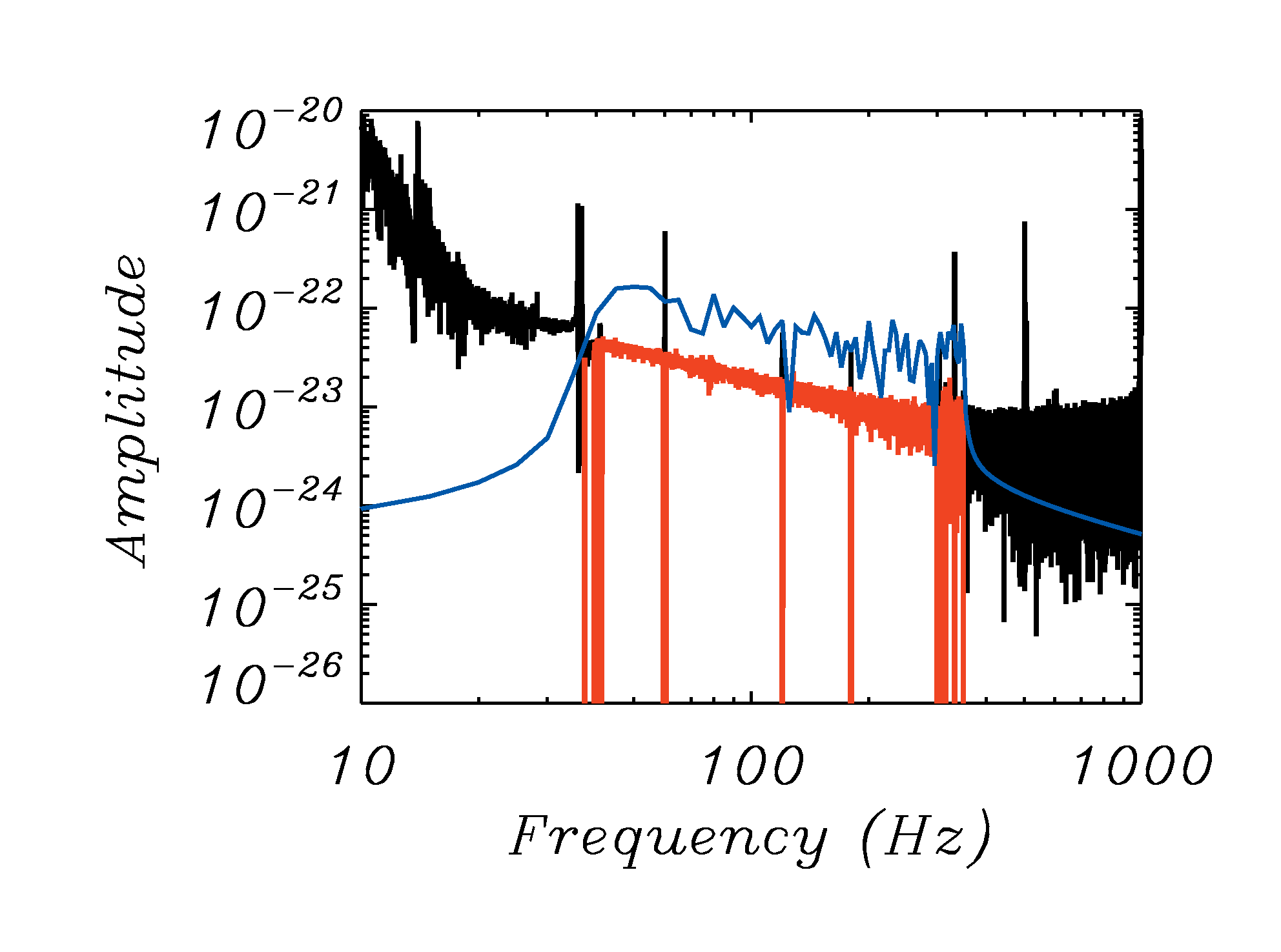}
\includegraphics[scale=0.14]{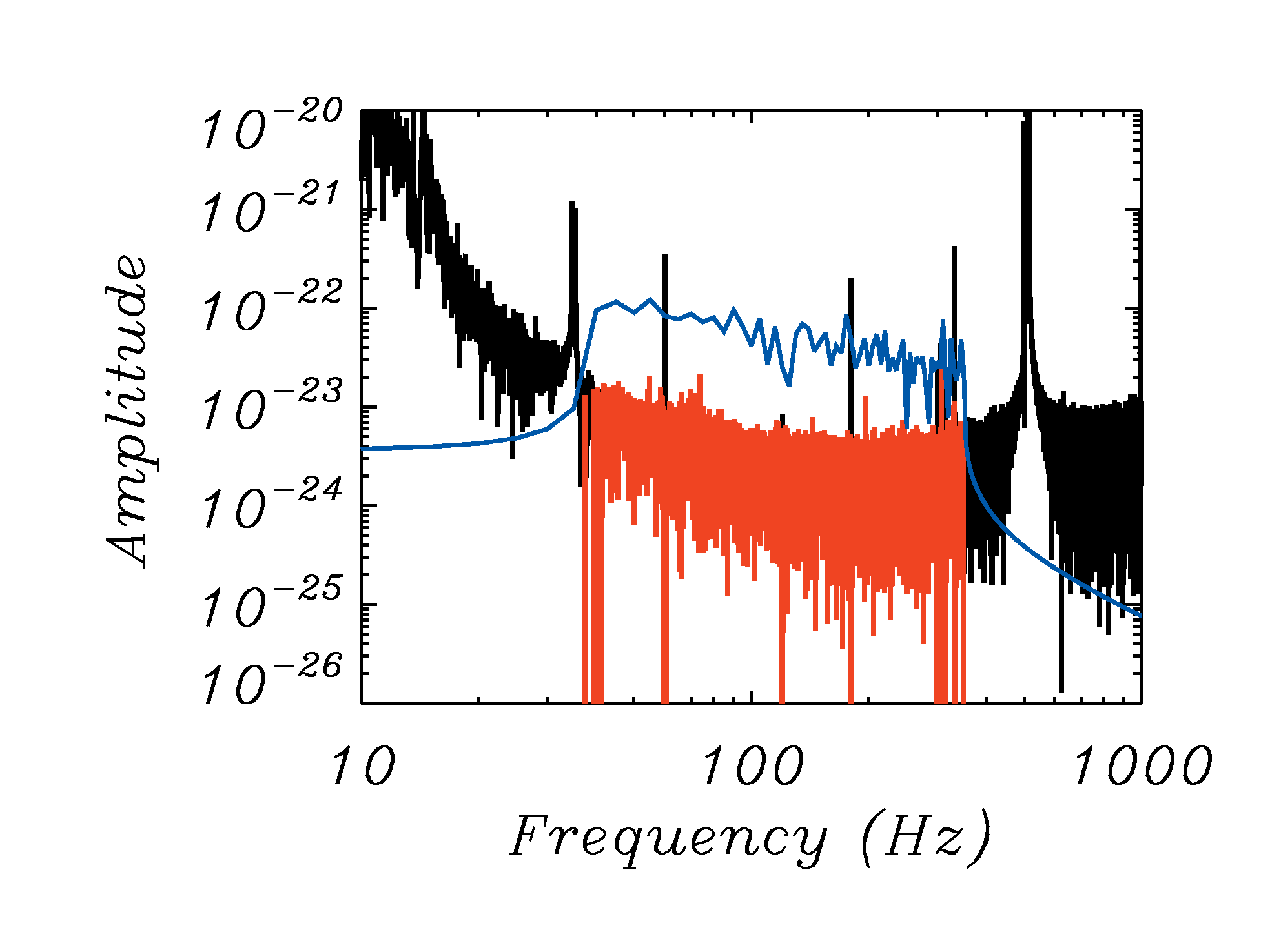}}}
\hbox{
 \centerline{\includegraphics[scale=0.13]{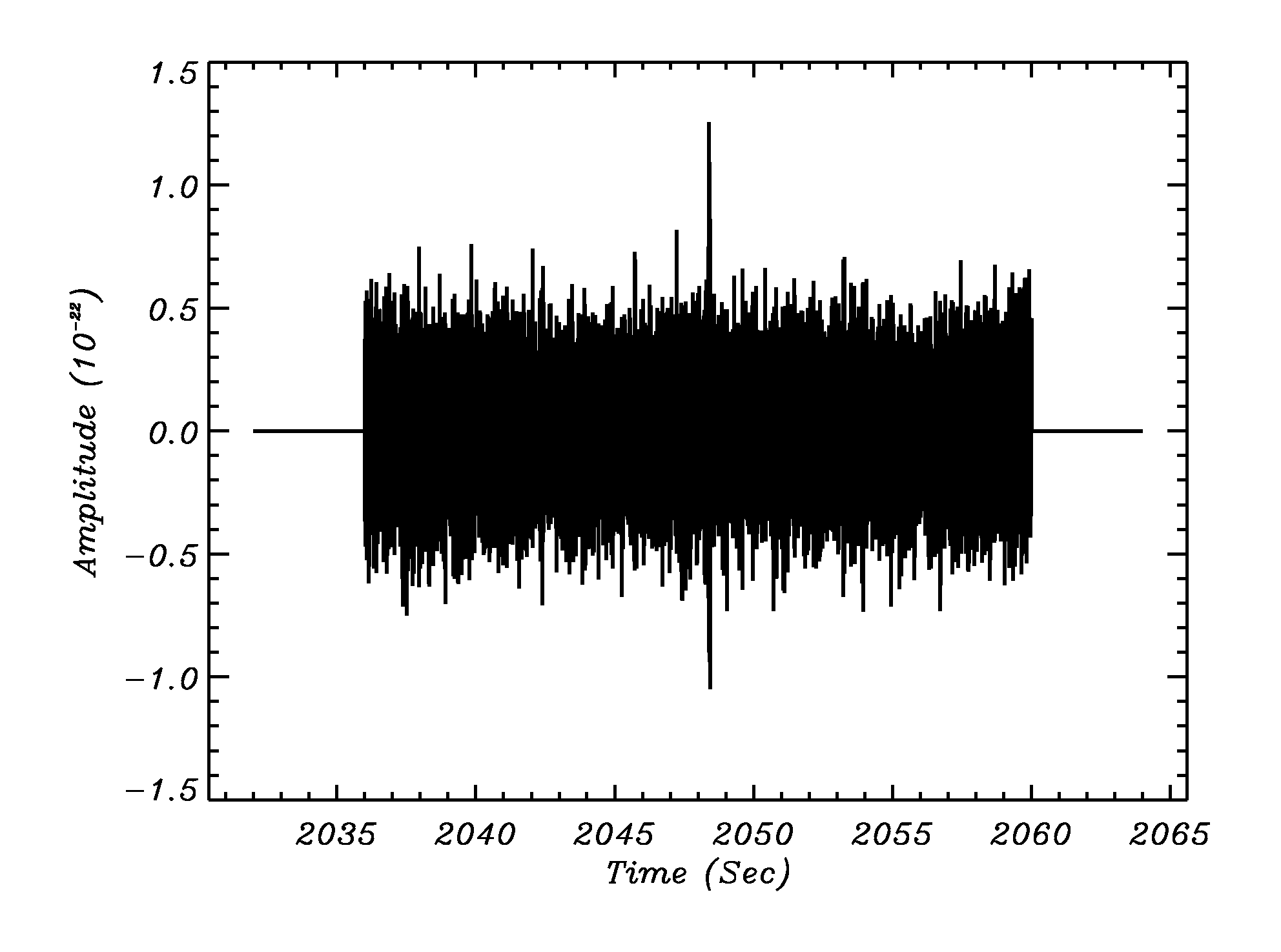}
\includegraphics[scale=0.13 ]{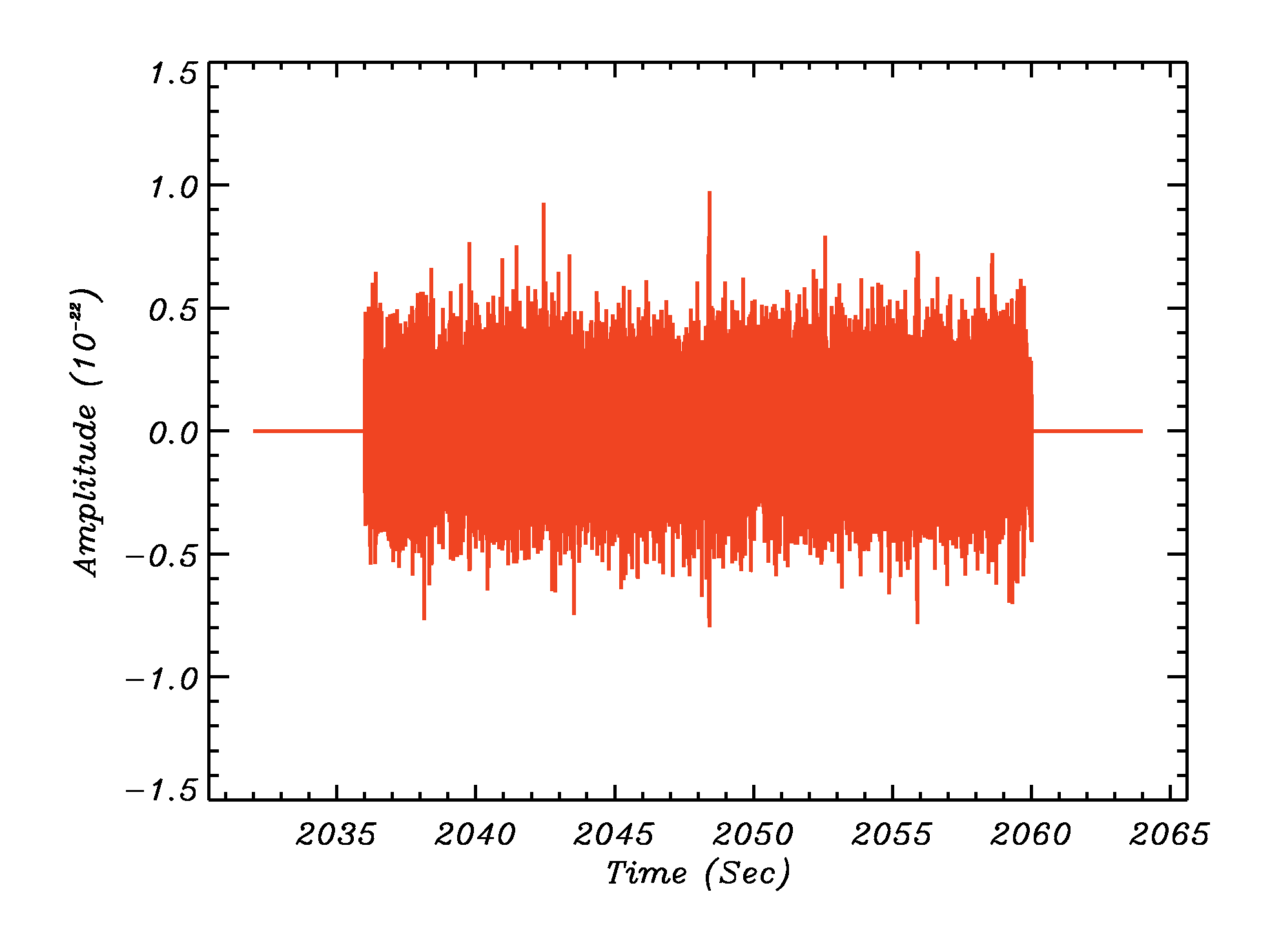}}}
 \caption{The top left panel shows the power spectrum of H32 before (black line) and after (red line) clipping. The top right
 panel shows the same for L32. The blue line corresponds to the power spectrum for the 0.2\,s GW150914 event
after clipping. (The enhancement of the amplitude for this record by a factor of $\sqrt{T_{32s}/T_{0.2s}}$ is due to
the different lengths of the H32, L32 and GW150914 records.) The bottom panels show the time records for H32 (left) and L32 (right) after clipping but without displacement and inversion.}
 \label{fig2}
 \end{center}
\end{figure*}
Fig.~\ref{fig2} shows the power spectra of H32 and L32 before clipping and after all narrow peaks mentioned above have been identified and removed (the vertical red lines) . The lower panel in Fig.~\ref{fig2} shows the dramatic reduction in the amplitudes of the filtered and cleaned records with respect to Fig.~\ref{fig1}. The peaks in the middle of these records indicate the position of GW150914 patches. We have verified that, except for these peaks, the resulting noise distributions are Gaussian and their variances are almost equal.

\begin{figure*}[!htb]
 \begin{center}
\hbox{
 \centerline{\includegraphics[scale=0.13]{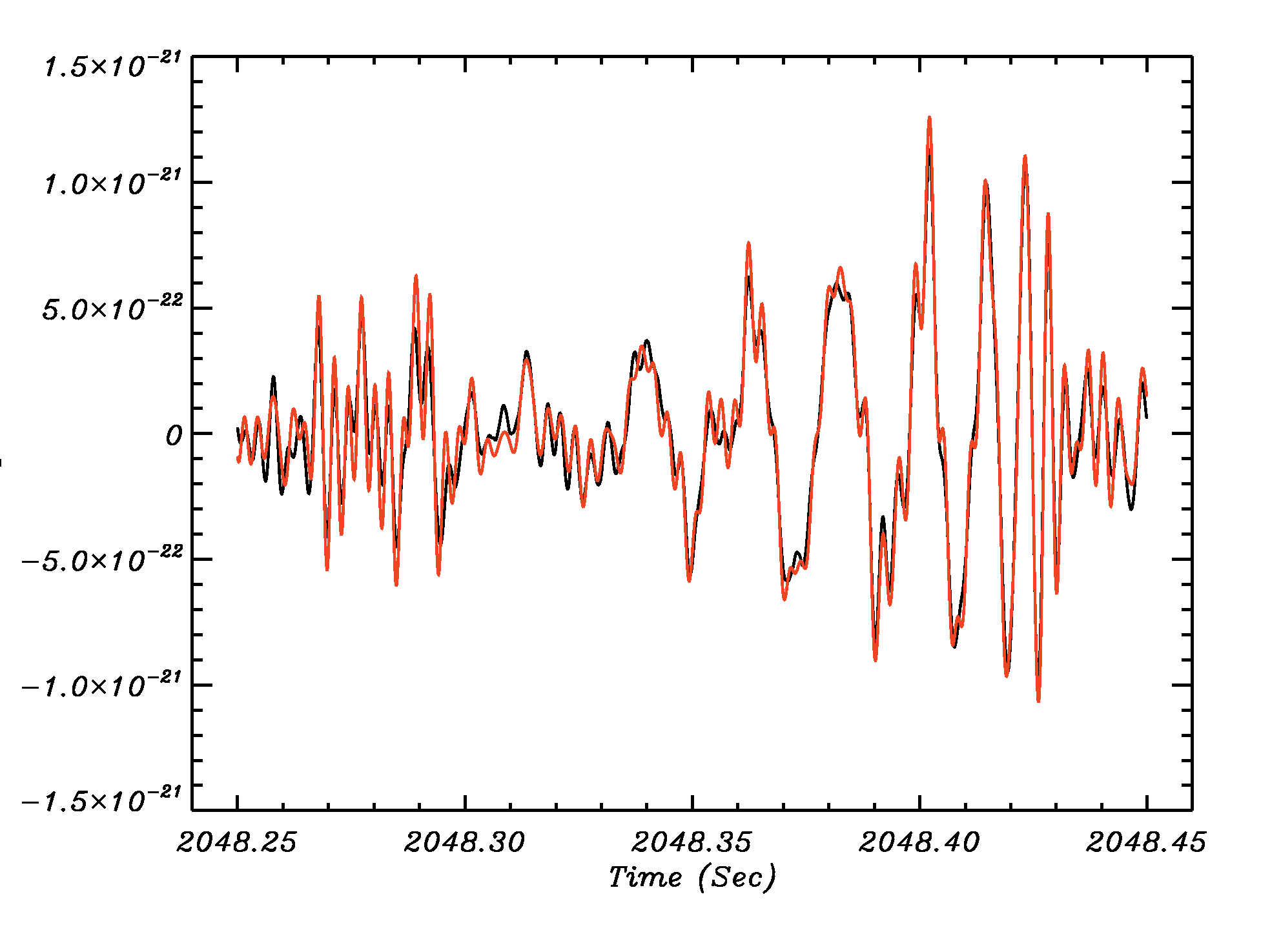}
\includegraphics[scale=0.13]{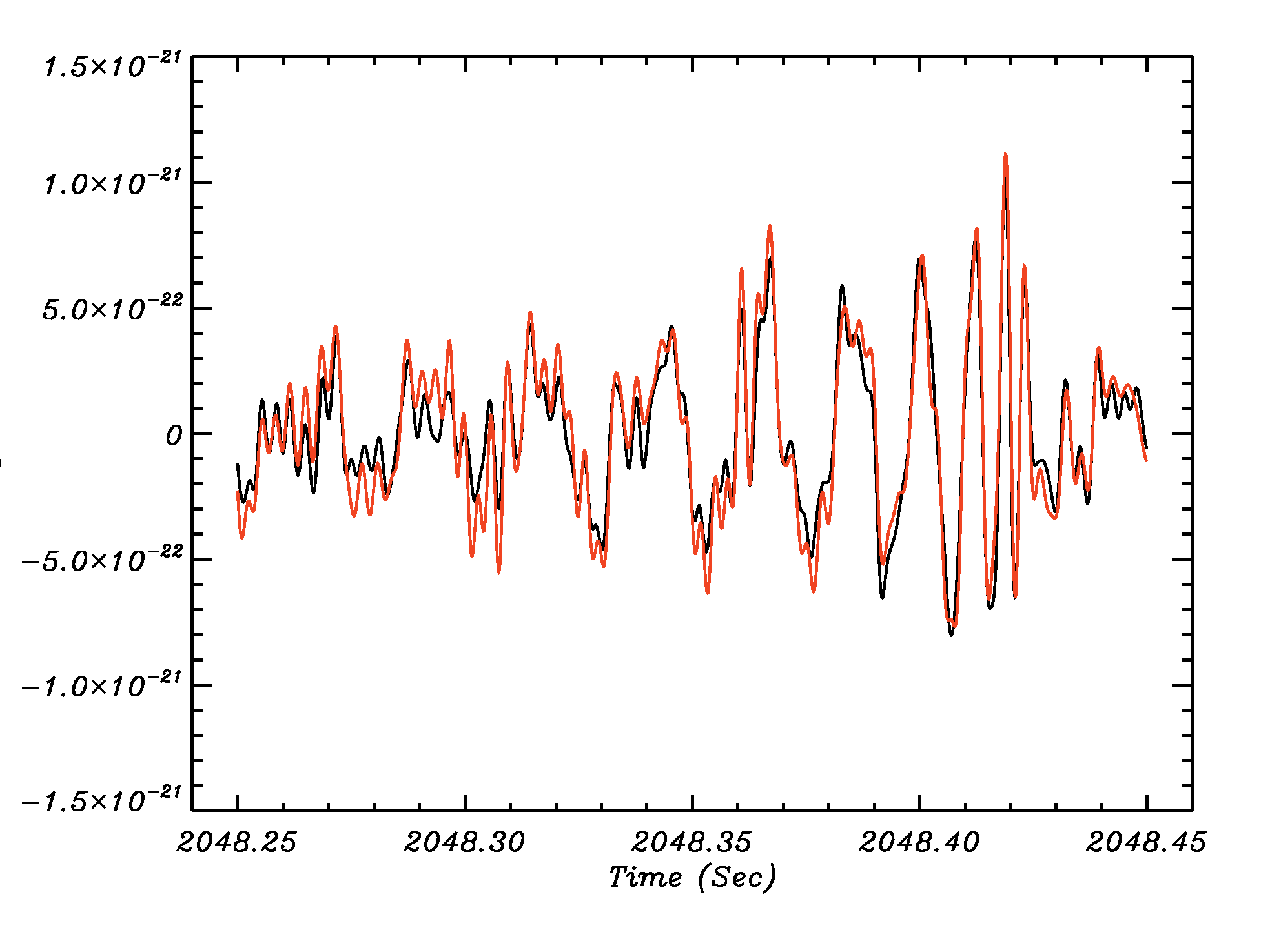}}}
\caption{Comparison of the reconstructed 0.2 second GW150914 event by LIGO (black) and by us (red). The agreement of these results over the entire time interval for both H (left) and L (right) is striking and provides evidence for the validity of our data processing.}
 \label{fig3}
 \end{center}
\end{figure*}
\begin{figure}[!htb]
 \begin{center}
\hbox{
 \centerline{\includegraphics[scale=0.12]{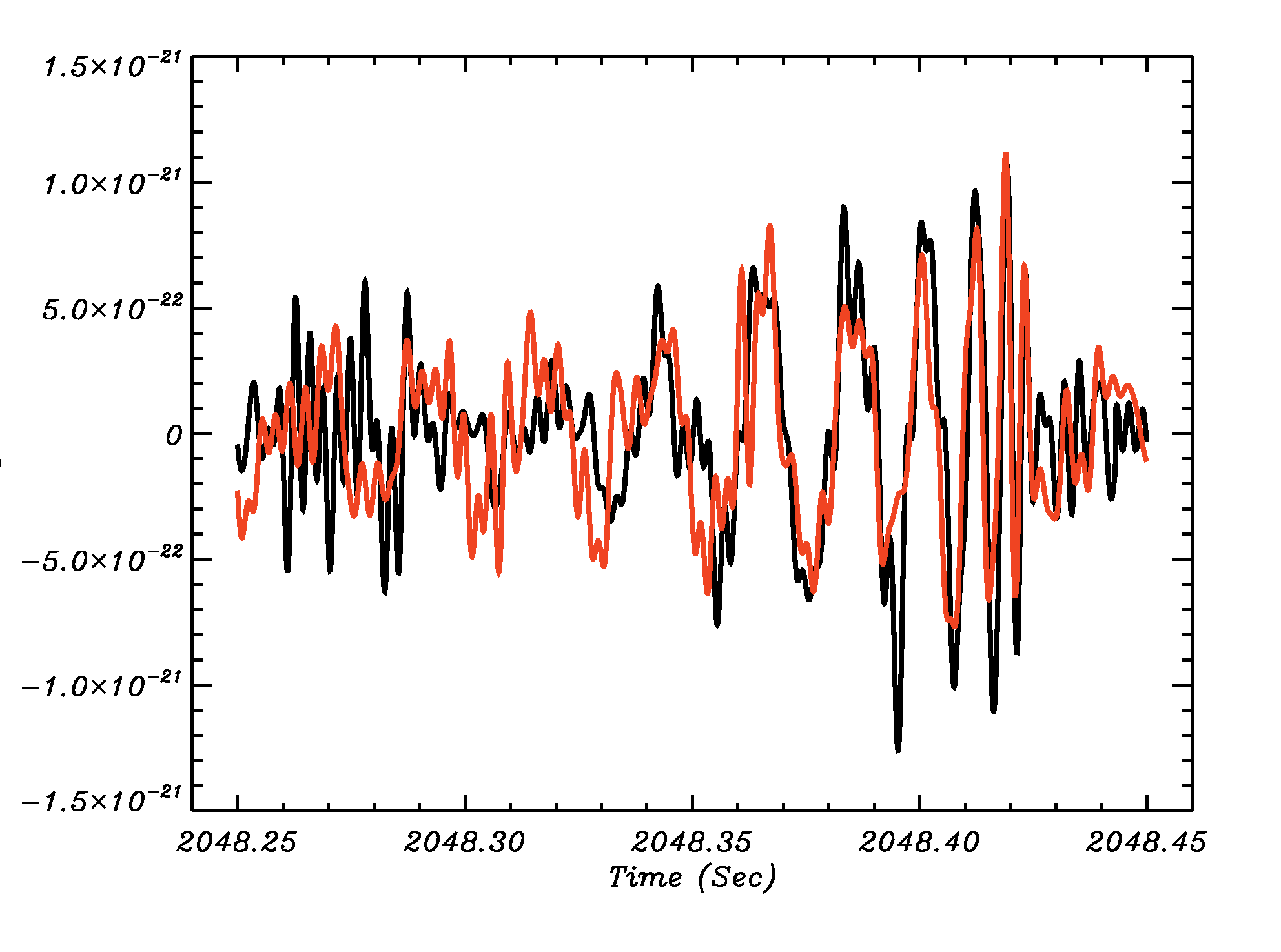}}}
 \caption{The 0.2 second H (black) and L (red) records with clipping and displacement for $\tau = 6.9$\,ms and inversion.}
 \label{fig4}
 \end{center}
\end{figure}

The identity of the 200\,ms GW150914 records extracted here and those obtained in~\citep{LIGO PRL} is confirmed by the
corresponding coefficients of cross-correlation $C_H\simeq 0.97$ and $C_L\simeq 0.95$. However, it is clearly seen from
Fig.~\ref{fig3}, that the latter parts of these records (the 100\,ms from 2048.35 and to 2048.45 seconds) are more strongly correlated than the full 200\,ms records. In this case, the improvement is not of much significance: $C_H\simeq 0.99$ and $C_L\simeq 0.98$. However, this difference is of greater importance for the cross-correlation comparison between H32 and L32. In Fig.~\ref{fig4} we show the comparison of our extracted signals for the optimal displacement $\tau=6.9$\,ms and with the inversion of the H record.
Here, the H-L cross-correlation coefficient for the complete 200\,ms record is $C^{200}_{H,L}\simeq 0.67 $. For the
second half of the record, $C^{100}_{H,L}\simeq 0.79 $. These numbers are the same as those obtained in~\citep{LIGO PRL}.

\section{The running window correlation of the GW150914 event for the 32 second record}
\label{sub:the_running_window_correlation_of_the_gw150914_event_along_the_32_second_record}
In this section we consider how the detectability of the GW150914 event depends on the displacement parameter $\tau$.
We first fix the position of the GW150914 event in the L32 record and decompose this complete record into 160 equal patches of length 0.2\,s. The H32 record (inverted and displaced by $\tau$) has been decomposed into the same 0.2\,s patches.
Eq(\ref{eq1}) is then used for each pair of corresponding patches in H32 and L32 for a number of values of $\tau$ including the optimal value $\tau=6.9$ms.
\begin{figure*}[!htb]
\includegraphics[width=0.49\textwidth]{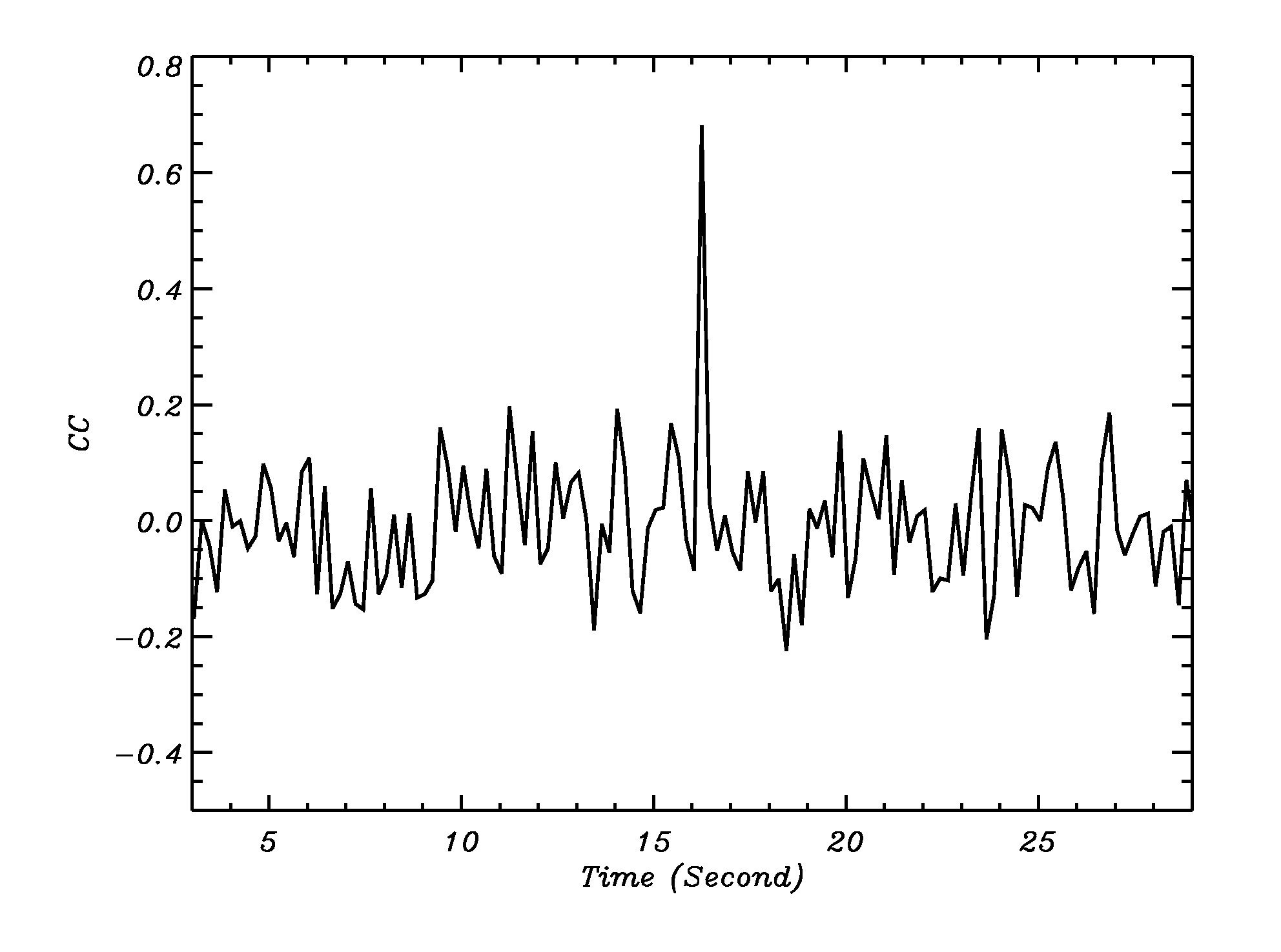}
\includegraphics[width=0.49\textwidth]{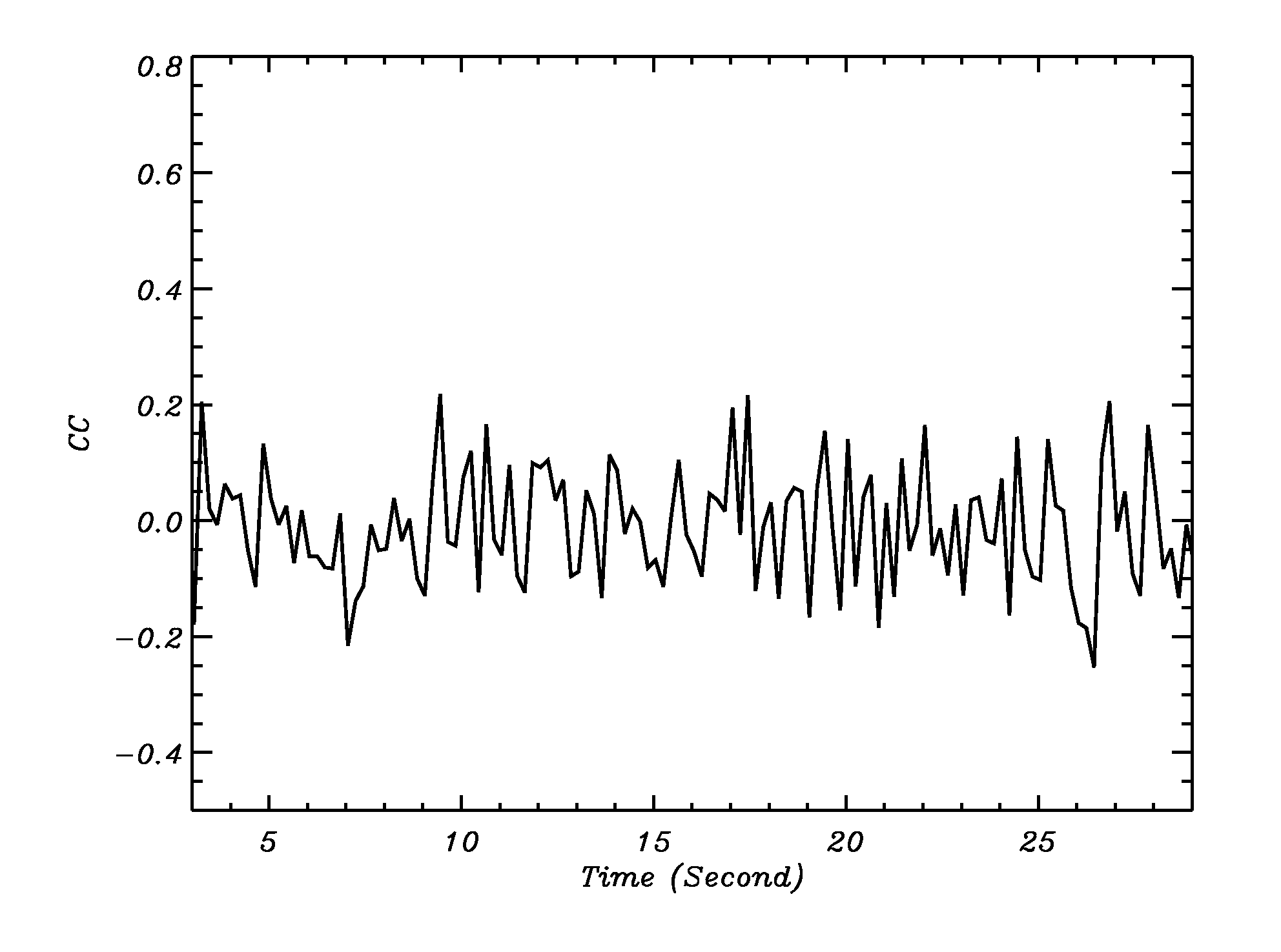}
\caption{The cross-correlation coefficients for H32 and L32 for $\tau=6.9$\,ms (left) and
$\tau=4.0$\,ms (right).}
 \label{tau}
\end{figure*}
Fig.~\ref{tau} shows the cross-correlation coefficients for $\tau=6.9$\,ms and for $\tau = 4$\,ms
(as representative of the correlations obtained for other values of the displacement $\tau$.) The peak in the
H-L correlator is readily identified for $\tau=6.9$\,ms; it is absent for $\tau=4$\,ms.

We note, however, that decomposition of the H32 and L32 records into 160 sub-records of length 0.2\,s could fail to identify genuine signals if a 0.2\,s GW event appeared partially in two neighbouring patches. To avoid this problem we adopt a ``running window'' approach to the calculation of the correlator. This approach is defined as follows: Let $G(t')$ be a short segment (e.g., length 0.1--0.2\,ms) of the GW150914 record from $t'=t'_{\rm on}$ to $t' = t'_{\rm off}$ and $G(t')=0$ otherwise. Let us consider any records, $y(t)$, with zero mean and a length greater than or equal to the length of $t'_{\rm off}-t'_{\rm on}$. The running window correlation with fixed $t$ is defined as:
\begin{eqnarray}
&& C(t)=\frac{1}{A}\sum_{t'=t'_{\rm on}}^{t'_{\rm off}}y(t+t')g(t'),\nonumber\\
&&A^2=\sum_{t'=t'_{\rm on}}^{t'_{\rm off}}g^2(t')
\sum_{t'=t'_{\rm on}}^{t'_{\rm off}}y^2(t+t'),\nonumber\\
&&g(t')=G(t')-\frac{1}{t'_{\rm off}-t'_{\rm on}}\sum_{t'=t'_{\rm on}}^{t'_{\rm off}}G(t')
\label{eq3}
\end{eqnarray}

\begin{figure*}[!htb]
 \begin{center}
\includegraphics[width=0.49\textwidth]{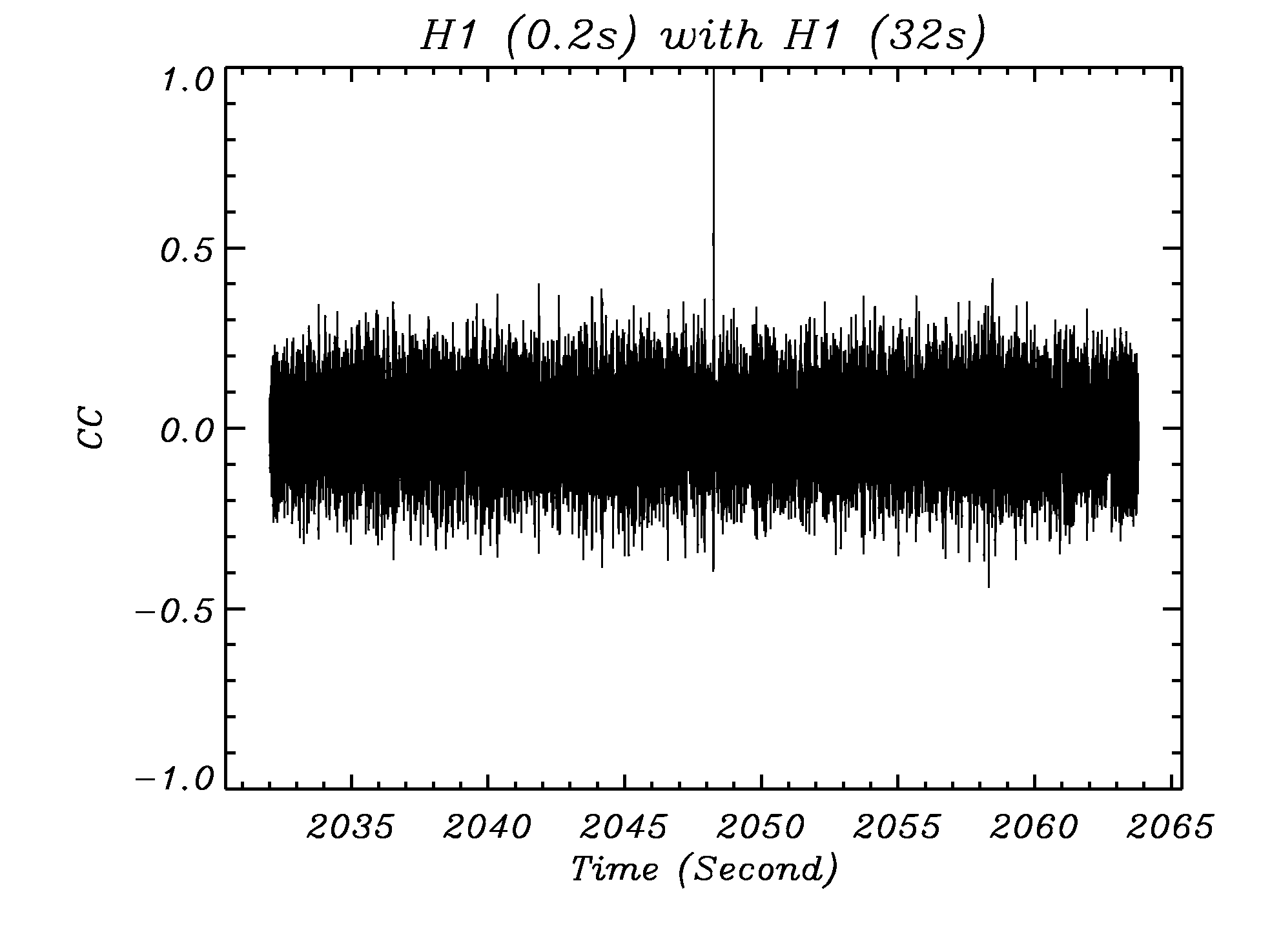}
\includegraphics[width=0.49\textwidth]{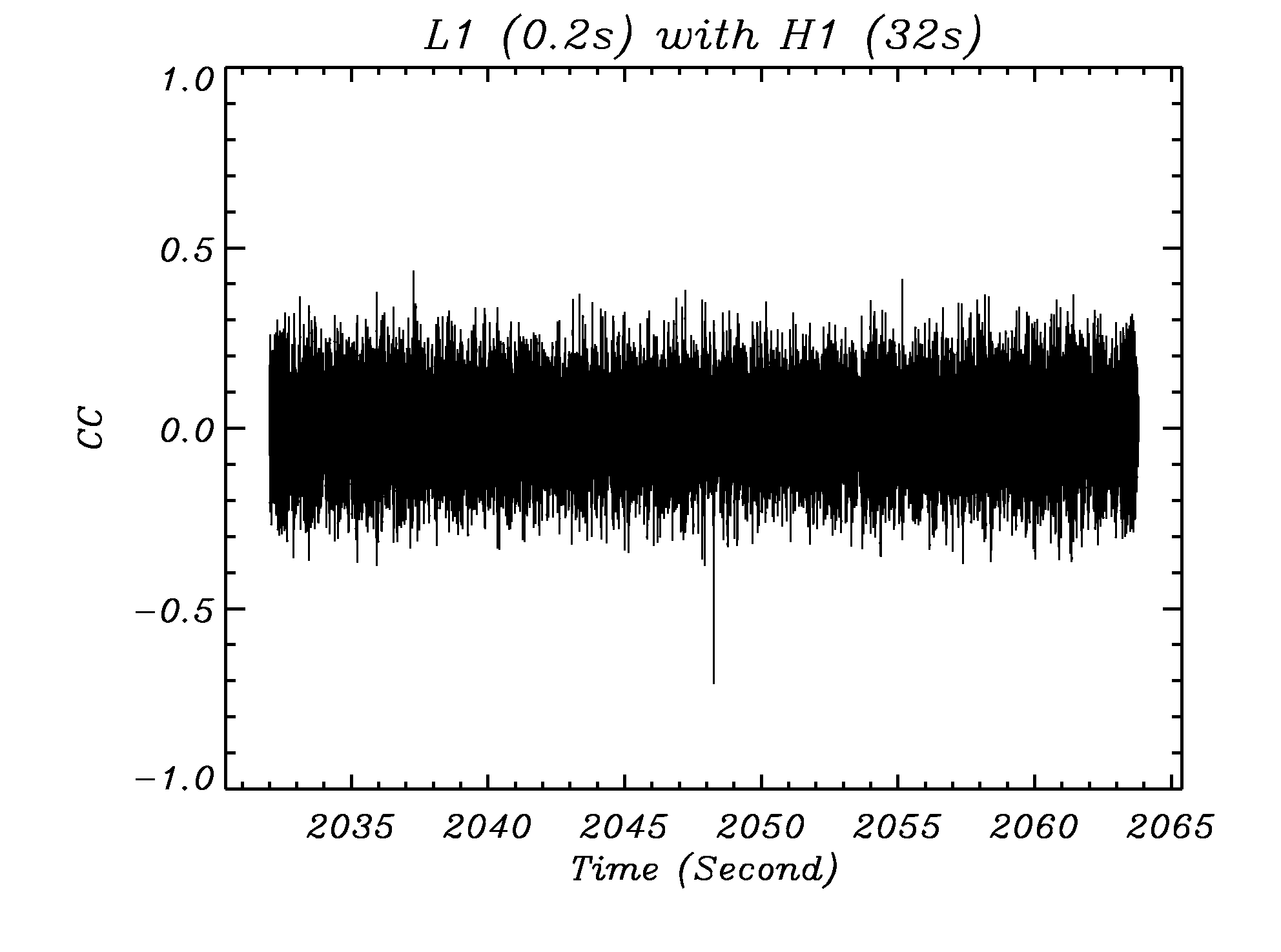}
\caption{The moving window cross-correlations for G(t) (0.2\,s) for the H32 record (left) and the L32 record (right).
Note that the signals have neither been displaced nor inverted.}
\label{fig4b}
\end{center}
\end{figure*}

The running window correlation is determined by moving the 0.2\,s window $G(t')$ bin-by-bin across
the H32 record and calculating the cross-correlation with the H32 component of the GW150914 event. The process is
repeated using the L32 record. The resulting cross-correlations are presented in
 Figure~\ref{fig4b}.

Fig.~\ref{fig4b} shows that the cross correlator with the H32 record has the expected value of $C=1$ at the position of the GW150914 event. The negative peak seen in the cross correlator with the L32 record indicates the need for the inversion of the GW150914 record in H32 and its shift by $\tau=6.9$ ms. The two methods (i.e., running and non-running cross-correlations) are complementary. The non-running calculation of the Pearson cross correlator offers a useful initial tool for performing a rapid
blind search for events of interest; the running method can be used for more accurate assessment once interesting candidates
have been found. Here, for example, the r.m.s value of the cross correlator for all the records with different $\tau$ (including the
$\tau=6.9$\,ms record) is Gaussian distributed with $\sigma=0.11$ for both the running and non-running cases, while the cross correlator for the GW150914 peak is $C^{200}_{H,L}\simeq 0.67$ (for the 0.2\,s case) and $C^{100}_{H,L}\simeq 0.79$ (for the 0.1\,s case). Thus, the signal at the optimal value of $\tau=6.9$\,ms is detected at the level $(6-7)\sigma$.

\section{The search for similar morphology in the 4096 second records}
The goal of this section is to extend our analysis to the far larger H4096 and L4096 records in order to search for events whose structure is similar to that of GW150914 but not present in the H32 and L32 TOD's. That fact that this record is 128 times longer than the 32\,s record considered so far will enable us to provide a better statistical description of the distribution of cross-correlations and will provide greater confidence that the GW150914 event is genuine. To be consistent with the LIGO procedure for the H4096 and L4096 TOD's, we have used
\begin{figure*}[!htb]
 \begin{center}
 \hbox{
 \centerline{\ \includegraphics[width=0.53\textwidth]{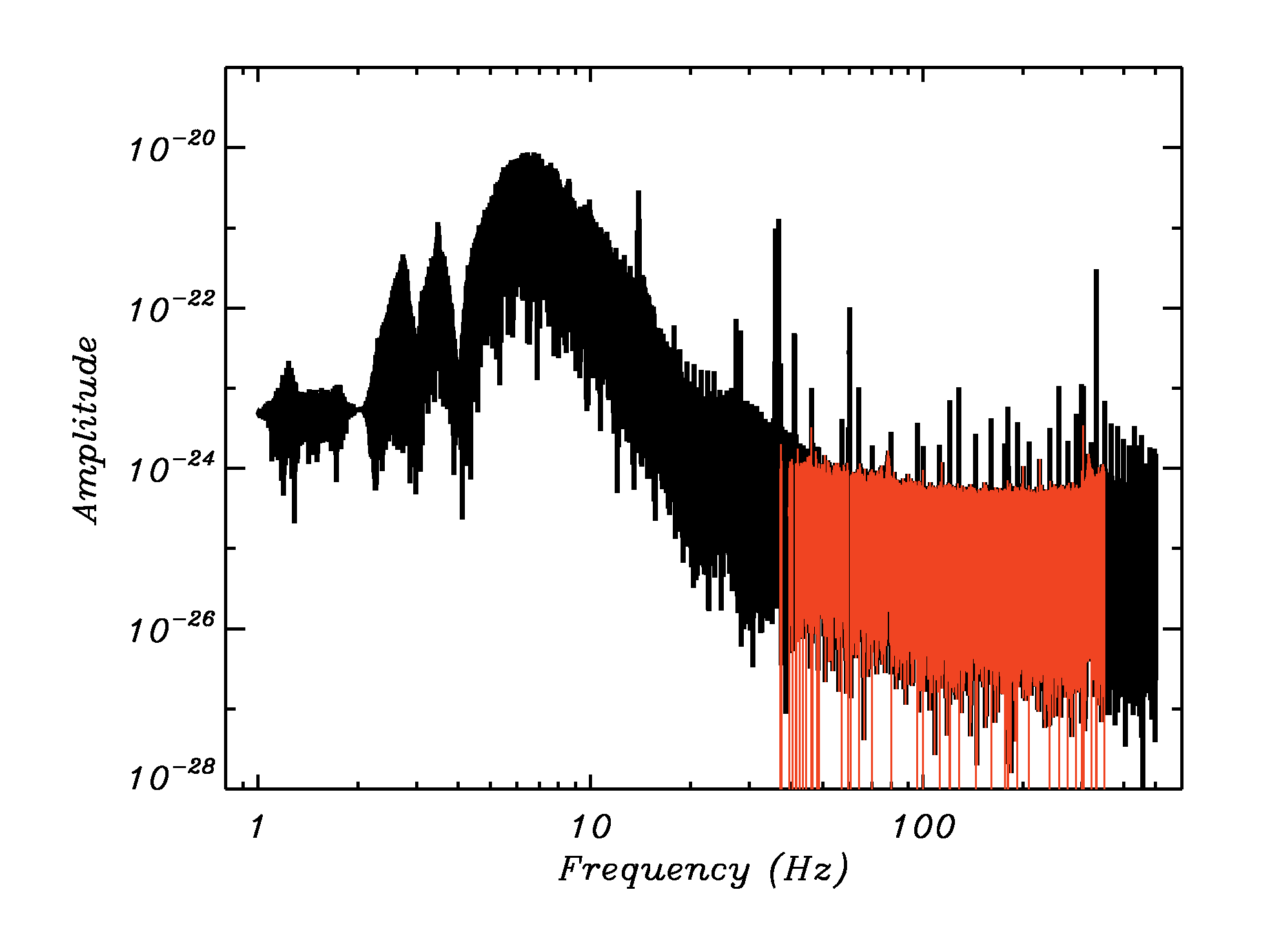}
 \includegraphics[width=0.53\textwidth]{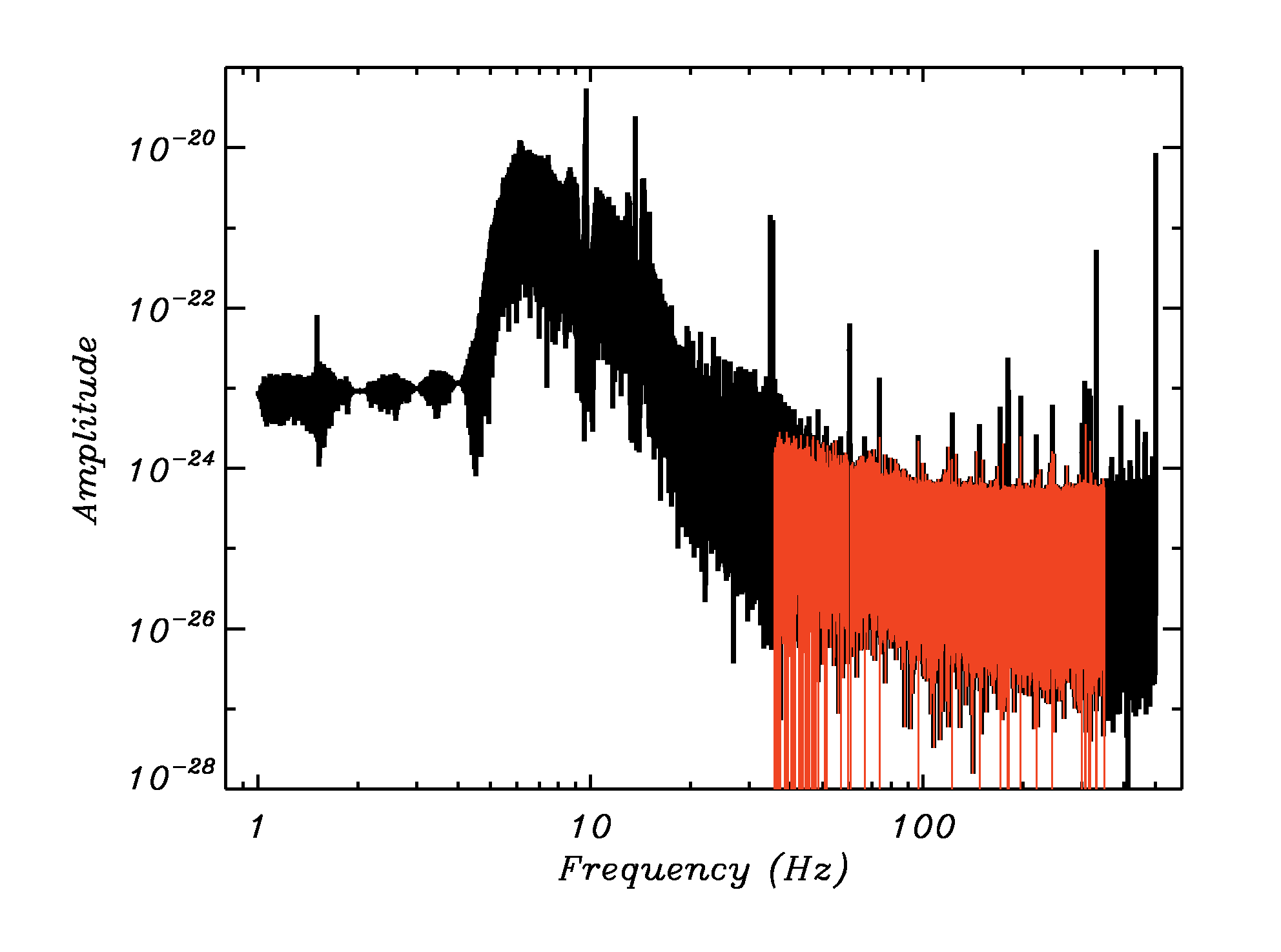}}}
 \caption{The amplitude of the power spectrum for the H4096 record (left) and the L4096 record (right).
The black lines show the unfiltered power spectra; the red lines show the result after filtering.
}
 \label{fig5}
 \end{center}
\end{figure*}
the Fourier space filtering from the publicly available LIGO ``strain-equivalent noise'' data, which is designated as $h(f)$ by the LIGO team in their paper \citep{LIGO Tech}. We first smooth the logarithm of $h(f)$ using a 50-bin running window.
We then define $d(f)$ as the difference between $\log(h(f))$ and the smoothed $\log(h(f))$. All frequencies with $d(f)$ higher than $4\sigma$ of $d(f)$ are removed from the analysis. The resulting power spectra for H4096 and L4096 are shown in Fig.~\ref{fig5}.

\begin{figure*}[!htb]
 \begin{center}
\hbox{
% \centerline{\includegraphics[scale=0.14]{ps_H0H1.eps}%}}
%\hbox{
 \centerline{\includegraphics[scale=0.14]{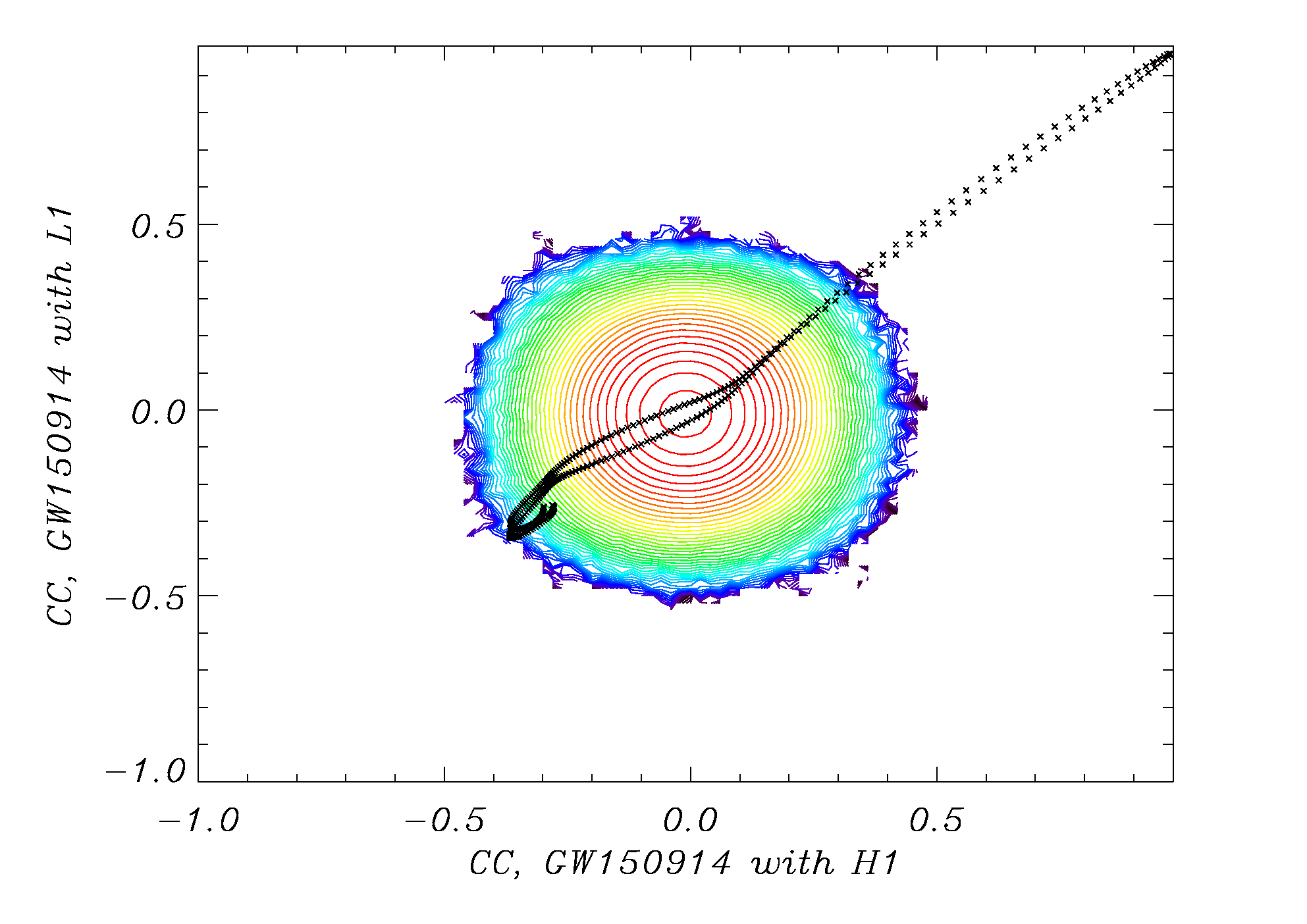}%}}
\includegraphics[scale=0.253]{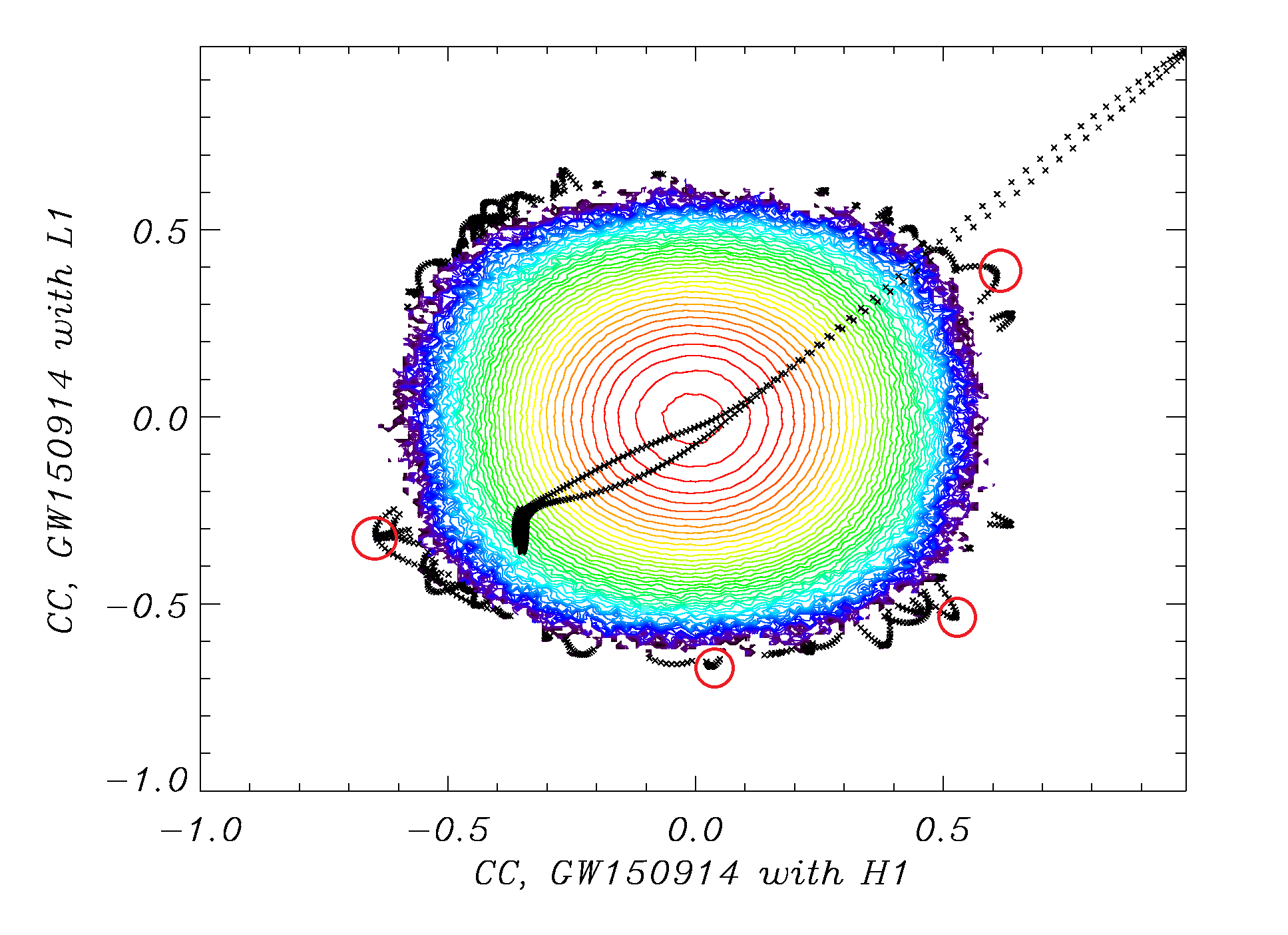}}}
 \caption{The running window cross-correlations for the GW150914 event using cleaned H4096 (H1) and L4096 (L1) records. The left panel corresponds to the GW150914 0.2\,s event. The right panel show the same cross-correlations for the second 0.1\,s portion of this event. The black lines correspond to the trajectory of cross-correlations between the GW150914 event as a function of the position of the running window. Exact overlap corresponds to the point $(x=1,y=1)$. The red circles indicate the positions of the A-D events, starting from that nearest to the black line peak (A) and proceeding clockwise to D.
}
 \label{fig6}
 \end{center}
\end{figure*}

After cleaning, we have used the running cross-correlation method to check for signals morphologically similar to the records,
GW150914$_{H,L}$, that characterize the GW150914 event. Specifically, we calculate the cross-correlation between each patch in
the H4096 data and GW150914$_{H}$ as well as the cross correlation between the corresponding patch in the L4096 data and
GW150914$_L$. These correlators serve as coordinates, $x=C($GW150914$_H$, H) and $y=C($GW150914$_L$,L)
for the two-dimensional plot shown in Fig.~\ref{fig6}. We have investigated two special cases. The first corresponds to the entire 0.2\,s GW150914$_{H}$ and GW150914$_{L}$ record. The second is based on the second half of the GW150914$_{H}$ and GW150914$_{L}$ of length 0.1\,s. We note that both the 32\,s and 4096\,s records are significantly
oversampled, with strains reported once every 0.06\,ms. Thus, the 4096\,ms record contains some $6.8 \times 10^7$ points. In
\begin{figure*}[!htb]
 \begin{center}
\hbox{
 \centerline{\includegraphics[scale=0.12]{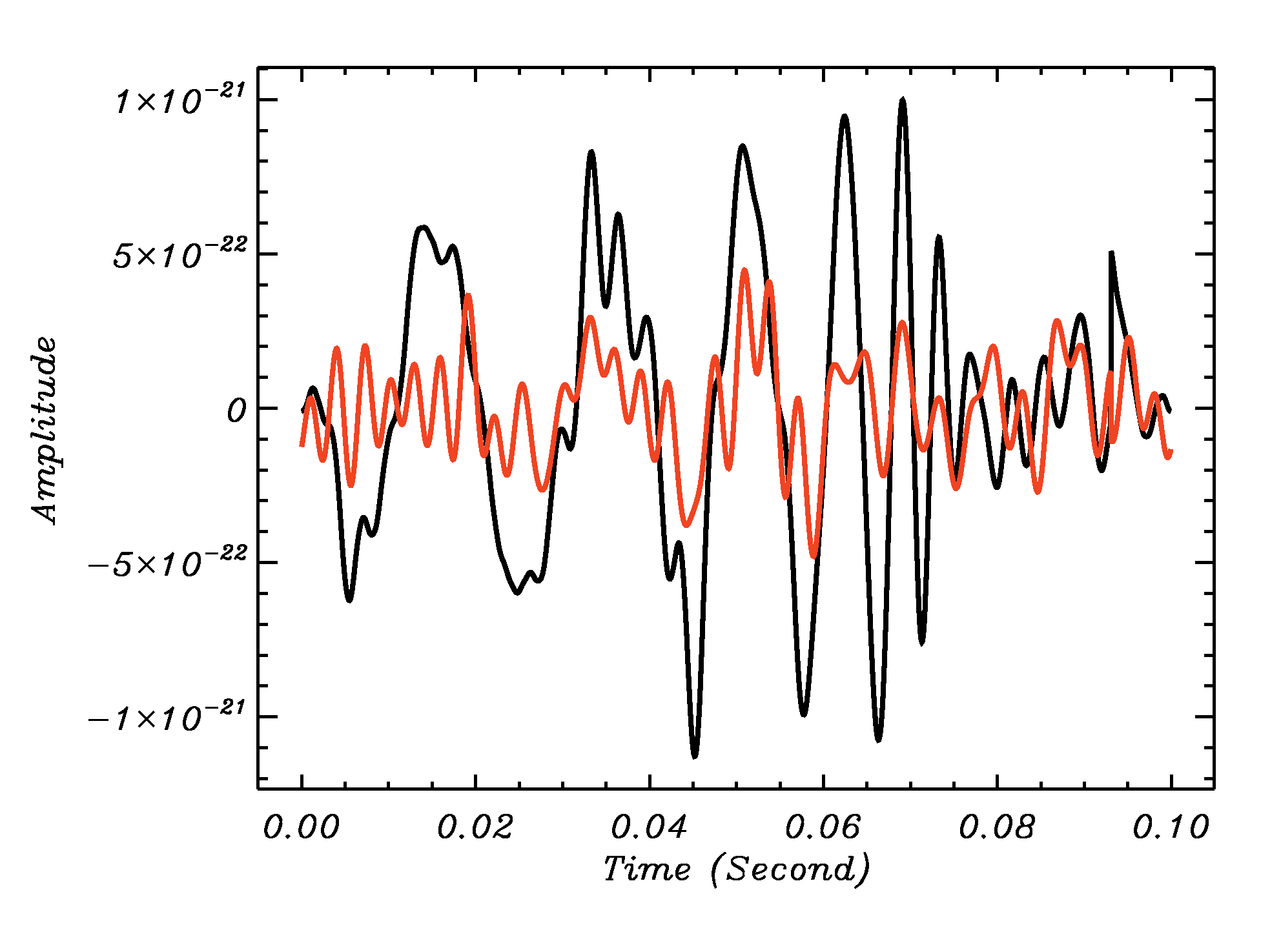}
\includegraphics[scale=0.12]{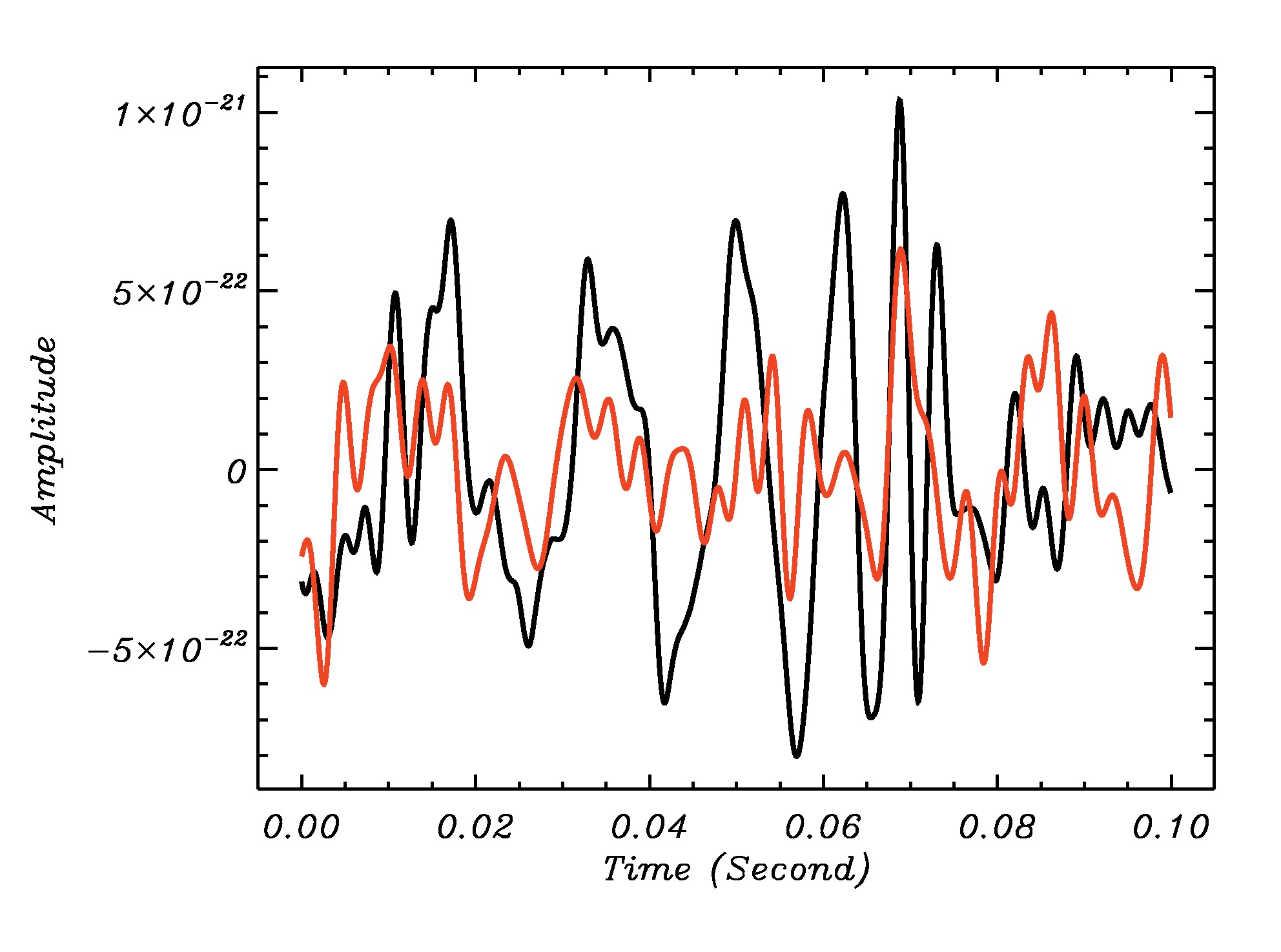}}}
\hbox{
 \centerline{\includegraphics[scale=0.12]{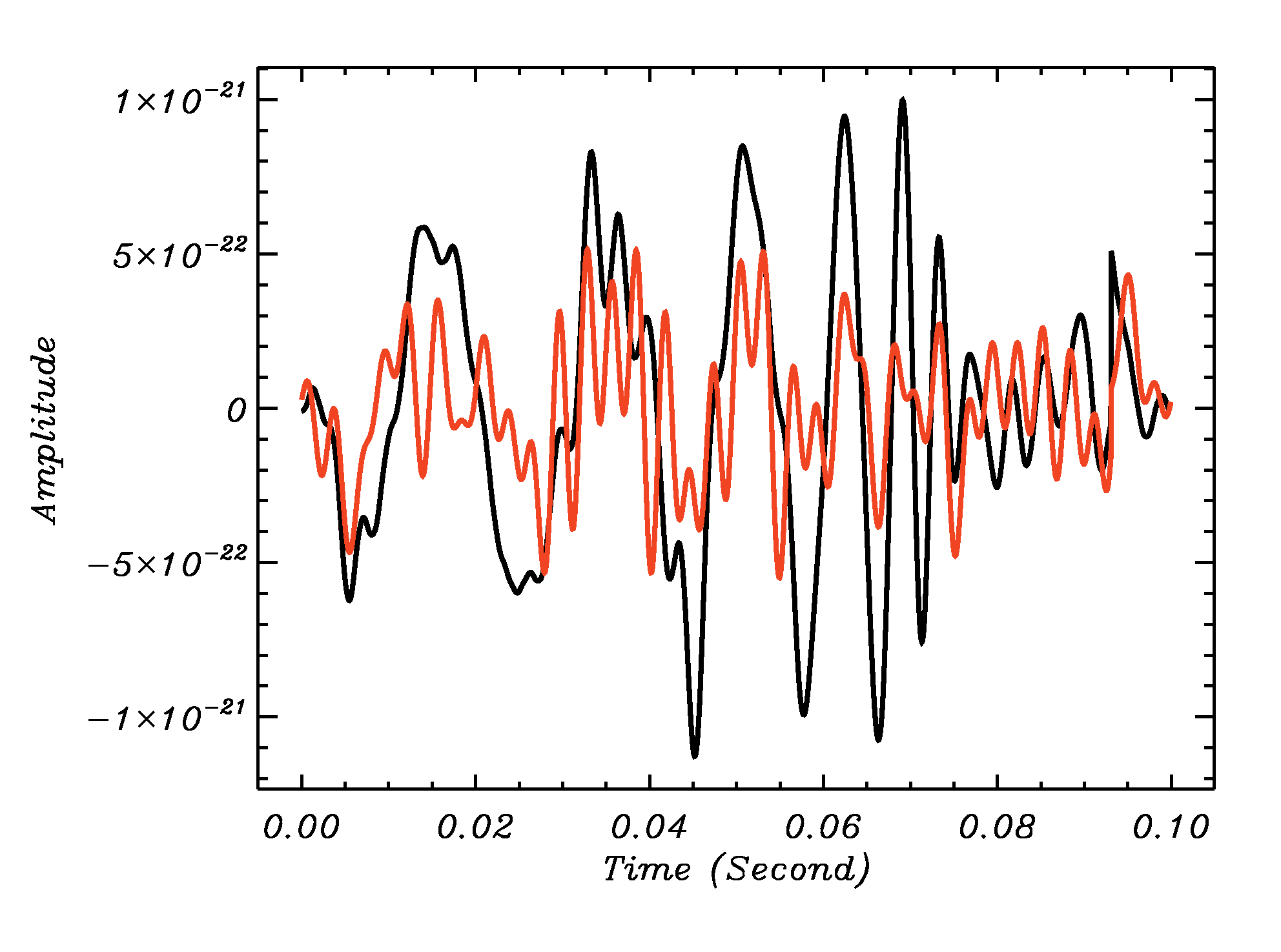}
\includegraphics[scale=0.12]{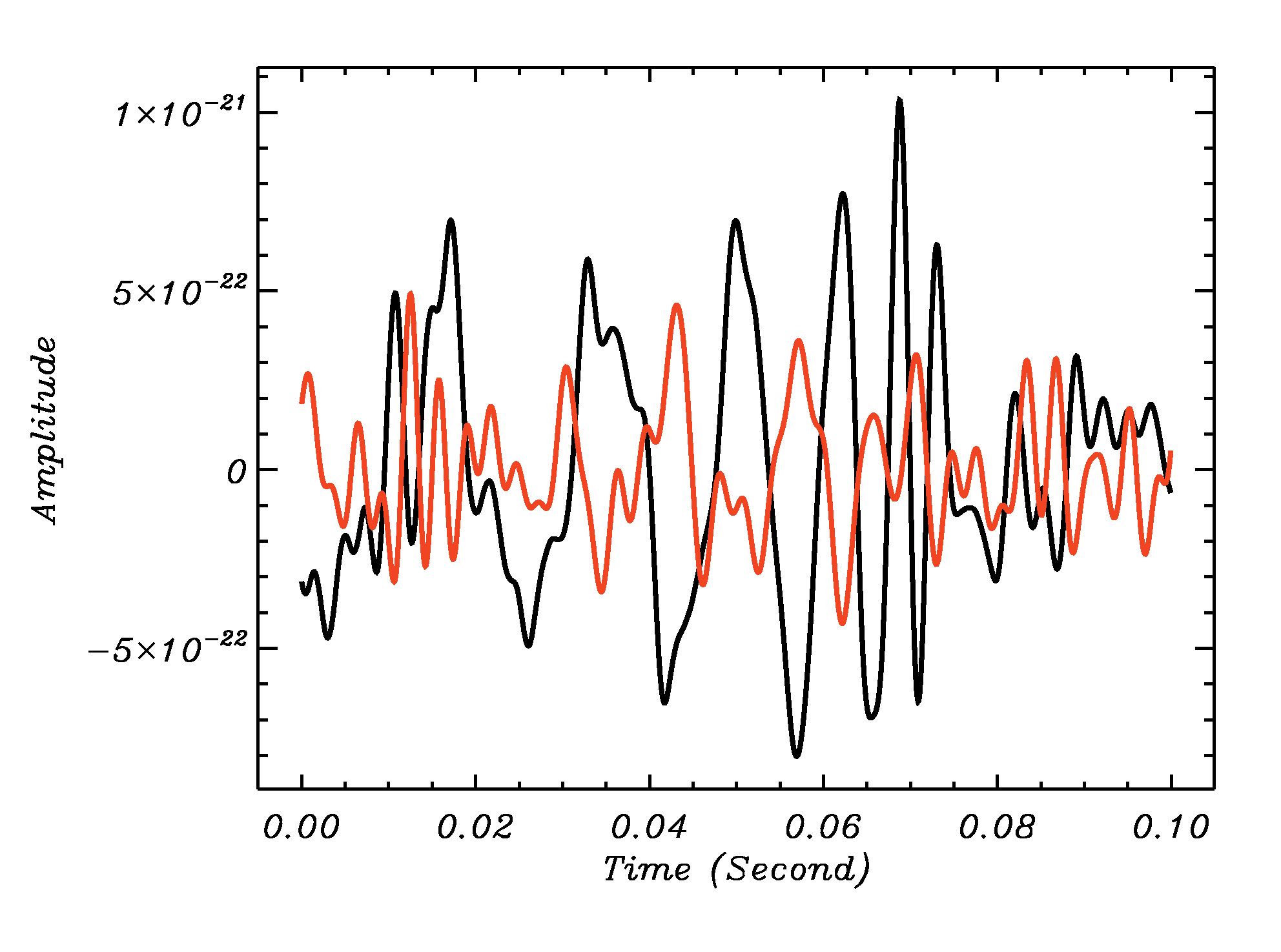}}}
\hbox{
 \centerline{\includegraphics[scale=0.12]{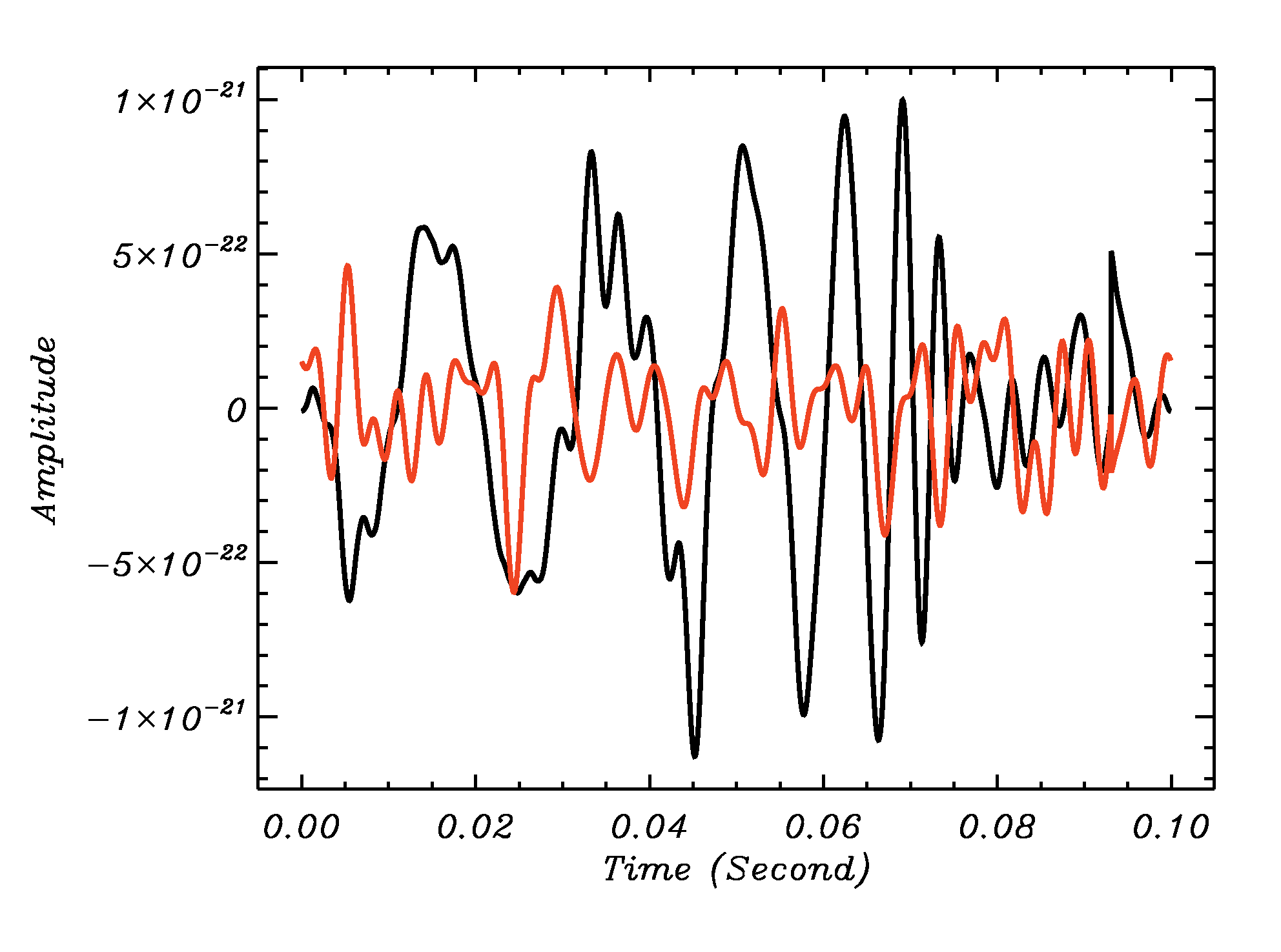}
\includegraphics[scale=0.12]{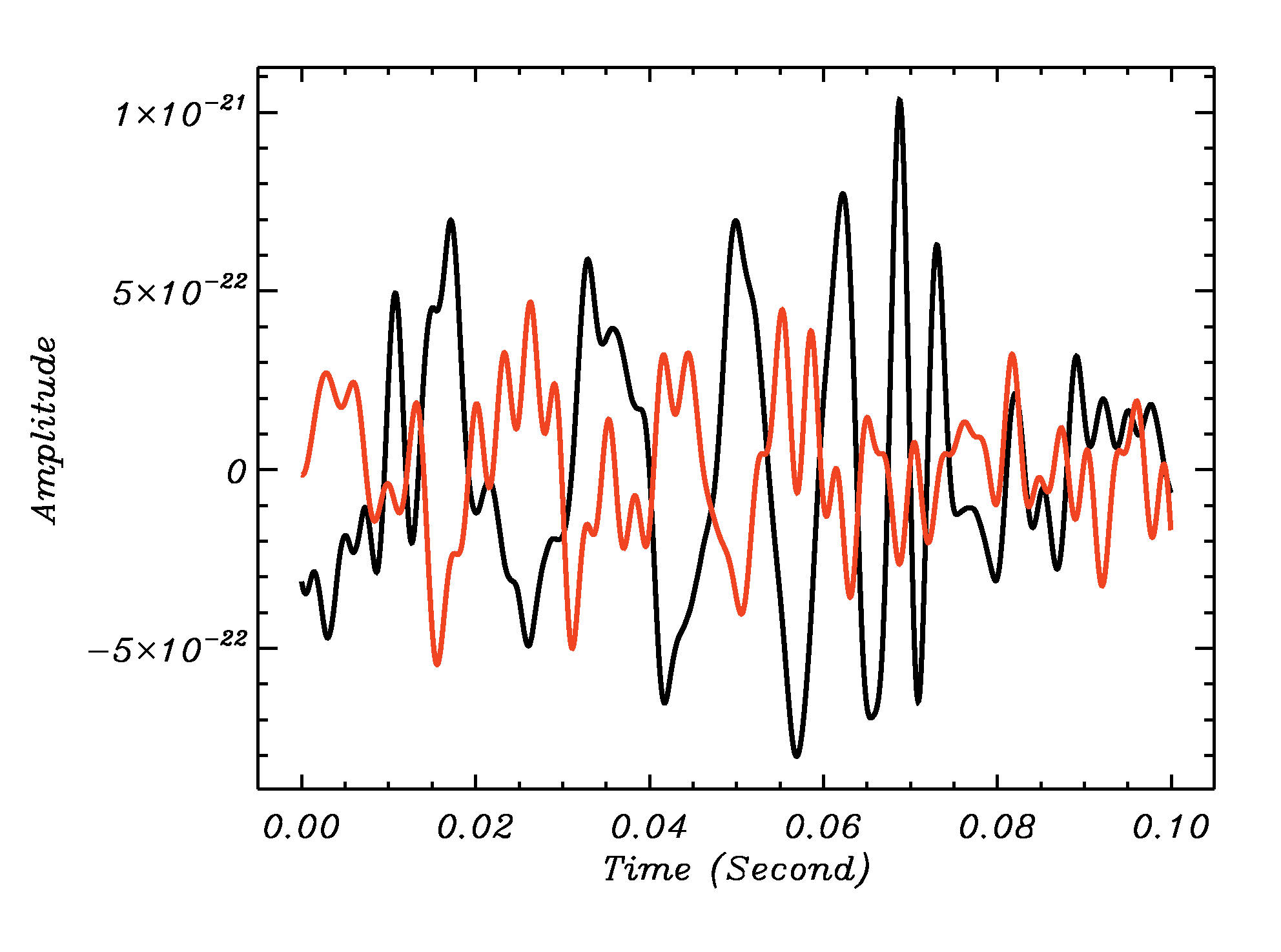}}}
\hbox{
 \centerline{\includegraphics[scale=0.12]{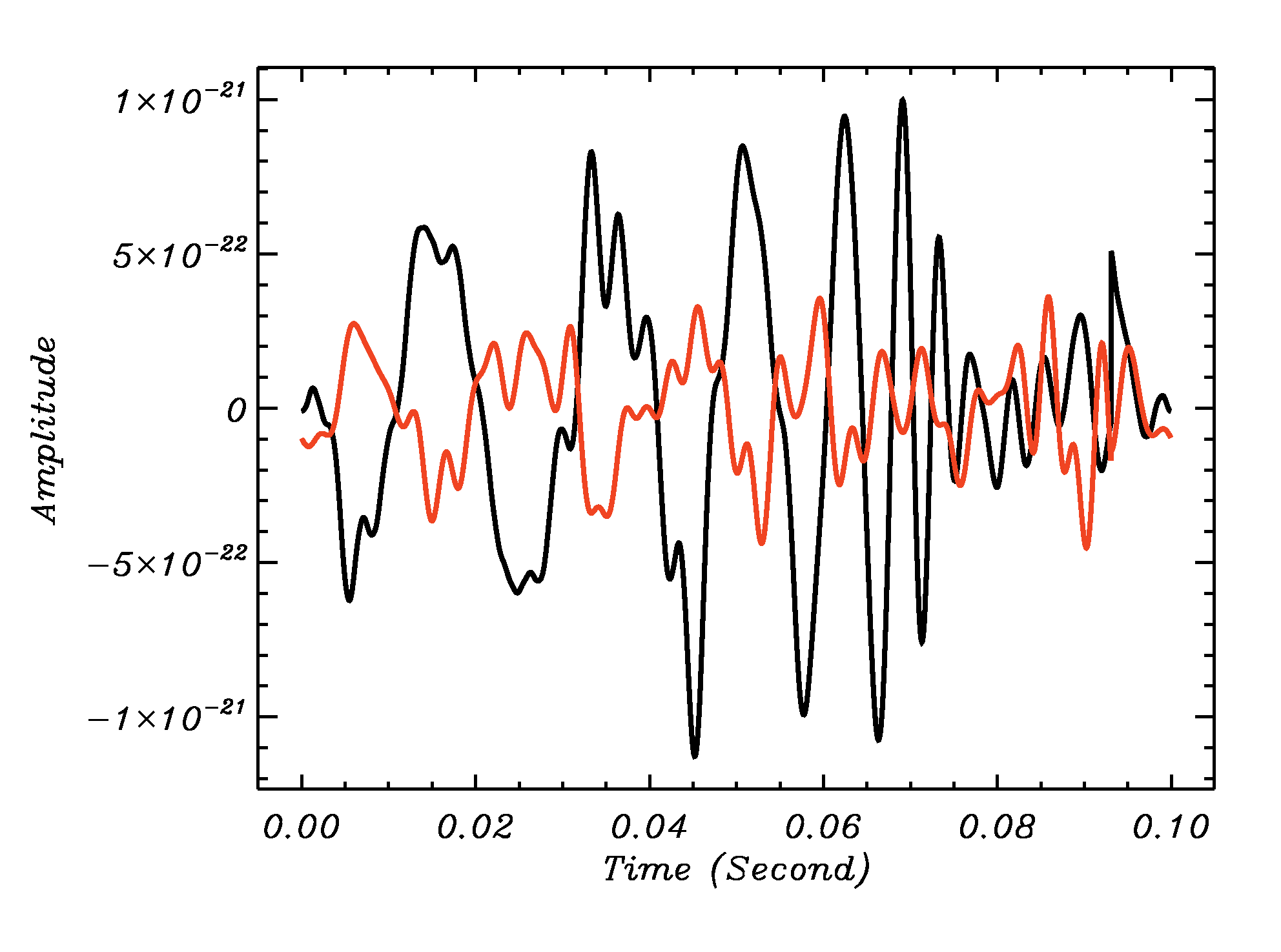}
\includegraphics[scale=0.12]{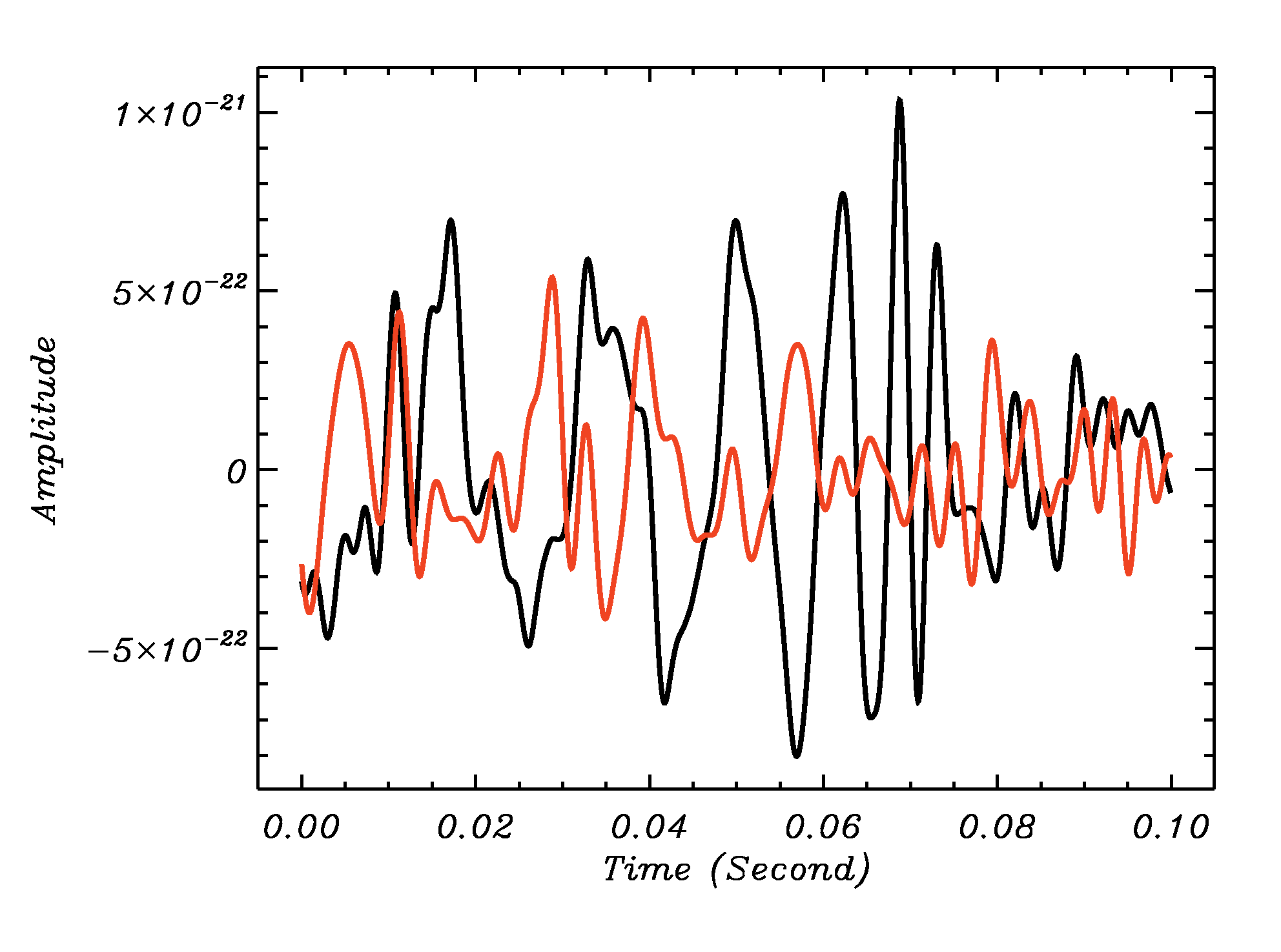}}}
\caption{From top to bottom: The A-D events of Fig.~8 (right panel) compared to the GW150914 event. H-H and L-L comparisons are shown in the left and right columns, respectively. %The thick black (for H) and red (for L) lines correspond to GW 150914 event after inversion and 6.9ms shift. The thin black and red lines correspond to the reconstructed events.
 }
 \label{fig7}
 \end{center}
\end{figure*}
Fig.~\ref{fig6} the contour lines from black to red colour reflect the number density of theses points. It is a clear consequence of
oversampling that these points will lie on two-dimensional trajectories. In particular, Fig.~\ref{fig6} shows one such trajectory
extending from the main cloud to the point $(1,1)$. This trajectory is the expected consequence of the window running
smoothly over the GW150914 event. It is clearly seen that the GW150914 event with the length 0.2\,s is genuinely unique. There
are, however, a few events that are morphologically similar to the GW150914 event for the 0.1\,s case with cross correlators at the
level of 0.3--0.6. In order to illustrate these features of the H4096 and L4096 records, we have considered the following four
events from Fig.~\ref{fig5} with the approximate coordinates $A=(0.6,0.4)$, $B=(0.5,-0.5)$, $C=(0,-0.7)$ and $D=(-0.65,-0.3)$. All of these events are characterized by a relatively large high level of positive or negative cross-correlations with GW150914. Event A is pointed in the direction of GW150914 and seems to be morphologically close to it. Event B is correlated with H1 but anti-correlated with L1. Event C has a pronounced anti-correlation with L1, but has no apparent correlation with H1. Event D illustrates the case where both H and L patches are inverted with respect GW150914. Fig.~\ref{fig7} shows the records in the time domain for patches A--D along with the records for GW150914. As usual, the Hanford record has been inverted and shifted by
6.9\,ms. We emphasize that none of these events can be regarded as candidates for a genuine GW event.

\section{The GW151226 event}
\begin{figure}[!htb]
 \begin{center}
\includegraphics[width=0.45\textwidth]{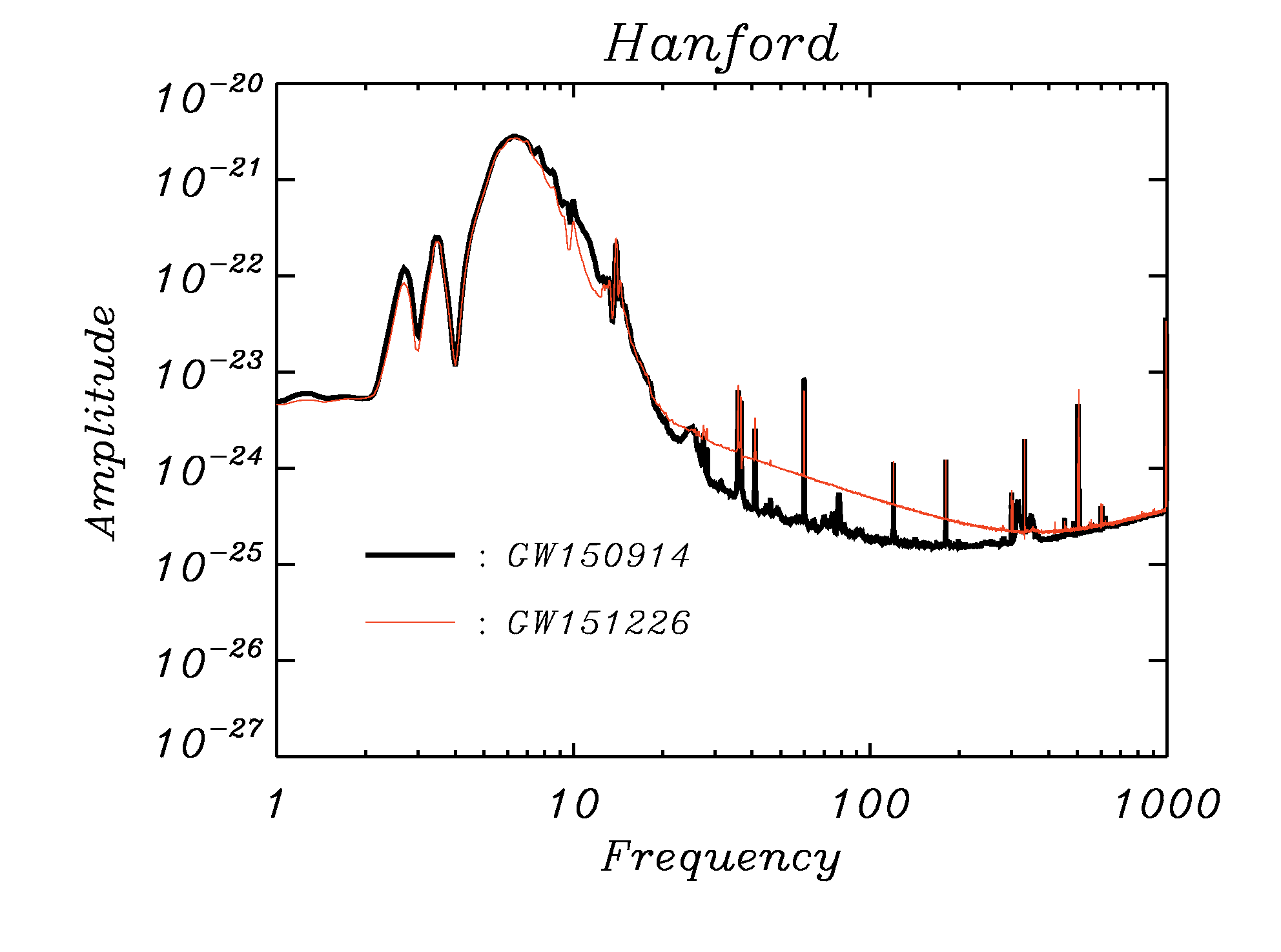}
\includegraphics[width=0.45\textwidth]{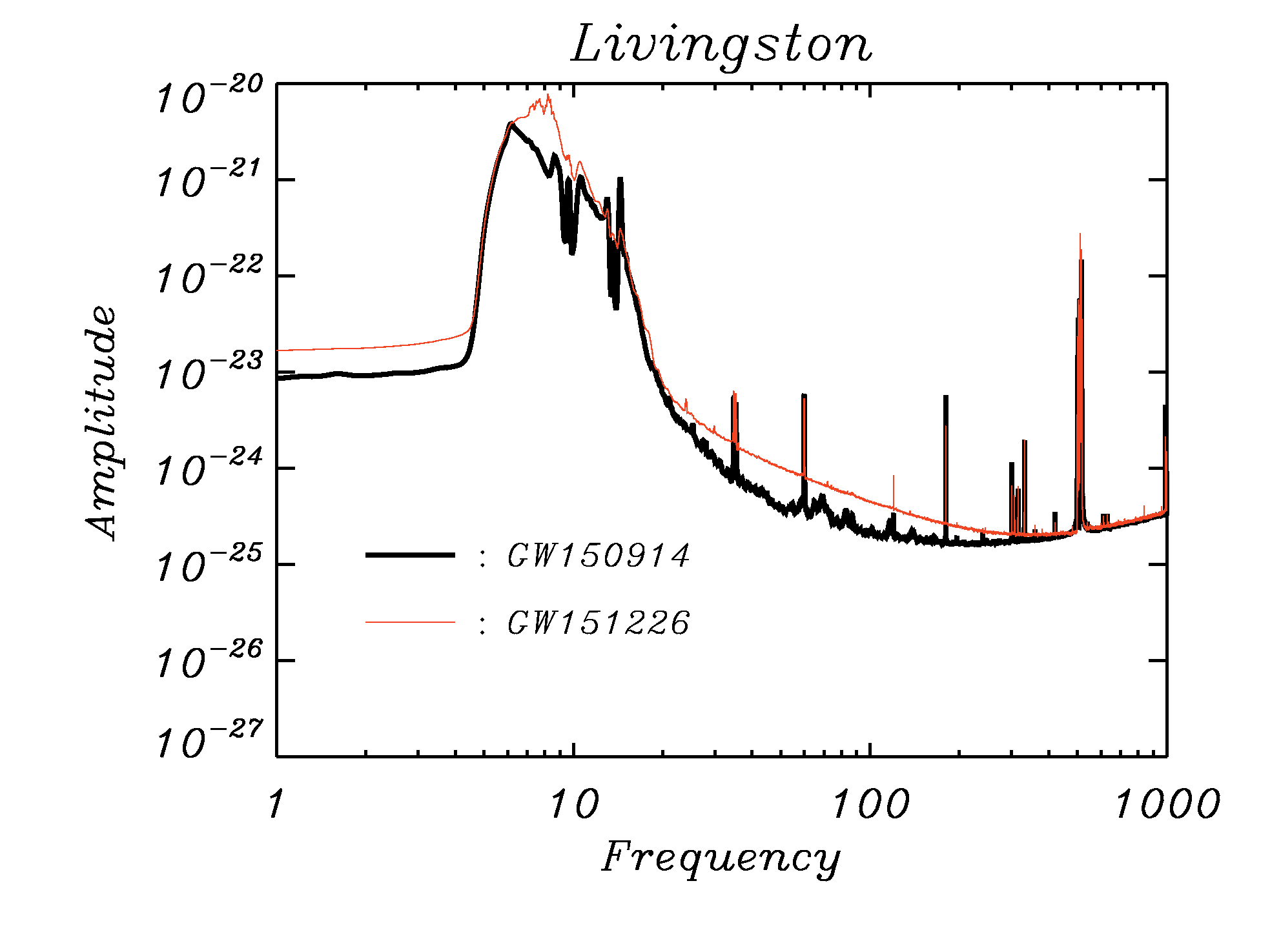}
\caption{Comparison of the noise spectrum derived from the 4096 second raw data for GW151226 (red) and GW150914 (black). The data has been smoothed by 0.25 Hz for clarity. }
 \label{fig:gw151226}
 \end{center}
\end{figure}

During preparation of this paper, the LIGO team released a second GW event designated as GW151226~\citep{LIGO GW2}.
This event has a peak gravitational strain that is roughly 3 times smaller than that of GW150914. Its frequency increases from 35 to 450\,Hz over a 2\,s time interval. (For comparison, the frequency of GW150914 increased from 35 to 250\,Hz over a 200\,ms interval.) The significance of GW151226 is claimed to be ``greater than 5$\sigma$'', which is comparable to the significance for GW150914 claimed in \citep{LIGO PRL} and confirmed here. Fortunately, the data for GW151226 is also publicly
available~\citep{LIGOData2}. We have repeated the present model-free analysis for this data set with appropriate adjustments of both the clipping frequencies and the time interval considered for the signal. Unfortunately, our template-free approach shows no statistically significant indication of {\em any\/} common signal in this case as a consequence of a less favorable signal-to-noise ratio. As noted, the GW151226 signal is materially weaker than that of GW150914. In addition, the noise level associated with
GW151226 is significantly larger as can be seen from Fig.~\ref{fig:gw151226}. This is likely to be due in part to the increased frequency range and signal duration claimed for GW151226.

As seen from Fig.~\ref{fig:gw151226}, the total power of the second LIGO data release significantly exceeds that of the first one (GW150914) in the frequency domain 25-300 Hz.  This enhancement of the power spectrum increases the contribution of noise to the cross-correlations and, consequently, decreases them. In this particular and probably more typical case of  noise realization, the implementation of templates for the GW-signal is essential.

\section{Discussion}

We have suggested a simplified method for the extraction of meaningful signals from LIGO data and have demonstrated
its ability to reproduce the LIGO collaboration's own results under the assumption that all narrow peaks in the power spectrum
can be ascribed to physically uninteresting signals. After clipping these peaks in frequency space and returning to the time domain, the GW150914 event is clearly seen above the broadband noise in the 32\,s records. Thus, the simple technique proposed here permits the identification of the GW150914 event without any assumption regarding its physical origin. We
have shown that the running cross-correlation method is a powerful tool for the quantitatively reliable extraction of this event
with minimal assumptions regarding the shape of the signal under investigation. We have confirmed that the LIGO GW150914 event is unique in the H and L detectors with an associated time displacement $\tau=6.9 \pm 0.4$\,ms at the level of 6--7$\sigma$, which is in agreement with LIGO's own analysis. At the same time we have identified a few others events in these records that are morphologically similar to GW150914 but with materially smaller cross-correlations with it. We have extended our analysis to
include the longer 4096 second records using the LIGO cleaning procedure. In spite of the increase statistical challenges posed by
this larger record, we have again confirmed the uniqueness of the GW150914 signal.

The strength (as well as the weakness) of our approach lies in the fact that it is almost entirely statistical. The only ``physical''
assumption is that all narrow resonances in the power spectrum are necessarily noise that can safely be eliminated independent of
their origin. We regard this method as both a supplement and a complement --- but certainly {\em not\/} an alternative --- to LIGO's
thorough analysis of the origins of noise in the power spectrum and its removal~\citep{LIGO PRL,LIGO Tech,LIGO TechA,LIGO TechB}. It is a supplement in the sense that it offers interested scientists an elementary tool that enables them to ``discover'' gravitational waves in the data that the LIGO collaboration has made publicly available. The conversion of a ``belief'' in LIGO's analysis to its ``understanding'' is of considerable value. Our method also offers a complement to the LIGO analysis. The fact that the analysis of ~\citep{LIGO PRL} involves binary black hole template forms might raise concerns that the extracted GW signals could be affected by theoretical bias. The agreement (i.e., the cross correlation) between the signals extracted by the
LIGO collaboration and by us (shown in Fig.~\ref{fig3} for both the Hanford and Livingston detectors) is impressive. Indeed, this agreement is materially stronger than the agreement between the Hanford and Livingston signals. This suggests that the use
of templates in \citep{LIGO PRL} has had little or no effect on the results obtained. Unfortunately, this is not always the case.

 In contrast to the unambiguous evidence for GW150914, our approach was unable to find any indication of the weaker signal
 of GW151226. In this case, the use of templates would appear to be essential. This suggests that the interpretation
 and even the existence of GW151226 can be affected by theoretical preconceptions. In such cases, independent observational evidence for the existence of an event would be highly desirable. At the very least, the meaning of the term ``significance'' would appear to be quite different in the two cases.

\section{Acknowledgments}

The authors would like to thank Slava Mukhanov, Alex Nielsen and Philipp Mertsch for valuable discussions and a critical reading of this paper. This research has made use of data and software obtained from the LIGO Open Science Center (https://losc.ligo.org), a service of LIGO Laboratory and the LIGO Scientific Collaboration. LIGO is funded by the U.S. National Science Foundation. This work was funded in part by the Danish National Research Foundation (DNRF) and by Villum Fonden through
the Deep Space project. Hao Liu is supported by the National Natural Science Foundation of China (Grant No. 11033003), the National Natural Science Foundation for Young Scientists of China (Grant No. 11203024) and the Youth Innovation Promotion Association, CAS.

\end{document}